\documentclass[
 onecolumn,
 amsmath,amssymb,
floatfix,superscriptaddress
]{revtex4-2}
\usepackage{graphicx}
\usepackage{dcolumn}
\usepackage{tikz}
\usepackage{placeins}
\usepackage{bm}
\begin{document}
\preprint{APS/123-QED}
\title{Instability and rupture of sheared viscous liquid nanofilms}
\author{Vira Dhaliwal}
\affiliation{Mechanics Division, Department of Mathematics, University of Oslo, 0316 Oslo, Norway}
\author{Christian Pedersen}
\affiliation{Mechanics Division, Department of Mathematics, University of Oslo, 0316 Oslo, Norway}
\author{Kheireddine Kadri}
\affiliation{Laboratoire PIMM, Arts et Métiers, CNRS, Hesam Université, 151 boulevard de l'Hopital, Paris, France}
\author{Guillaume Miquelard-Garnier}
\affiliation{Laboratoire PIMM, Arts et Métiers, CNRS, Hesam Université, 151 boulevard de l'Hopital, Paris, France}
\author{Cyrille Sollogoub}
\affiliation{Laboratoire PIMM, Arts et Métiers, CNRS, Hesam Université, 151 boulevard de l'Hopital, Paris, France}
\author{Jorge Peixinho}
\affiliation{Laboratoire PIMM, Arts et Métiers, CNRS, Hesam Université, 151 boulevard de l'Hopital, Paris, France}
\author{Thomas Salez}
\email{thomas.salez@cnrs.fr}
\affiliation{Univ. Bordeaux, CNRS, LOMA, UMR 5798, F-33405, Talence, France}
\author{Andreas Carlson}
\email{acarlson@math.uio.no}
\affiliation{Mechanics Division, Department of Mathematics, University of Oslo, 0316 Oslo, Norway}
\date{\today}
\begin{abstract}
Liquid nanofilms are ubiquitous in nature and technology, and their equilibrium and out-of-equilibrium dynamics are key to a multitude of phenomena and processes. We numerically study the evolution and rupture of viscous nanometric films, incorporating the effects of surface tension, van der waals forces, thermal fluctuations and viscous shear. We show that thermal fluctuations create perturbations that can trigger film rupture, but they do not significantly affect the growth rate of the perturbations. The  film rupture time can be predicted from a linear stability analysis of the governing thin film equation, by considering the most unstable wavelength and the thermal roughness. Furthermore, applying a sufficiently large unidirectional shear can stabilise large perturbations, creating a finite-amplitude travelling wave instead of film rupture. In contrast, in three dimensions, unidirectional shear does not inhibit rupture, as perturbations are not suppressed in the direction perpendicular to the applied shear. However, if the direction of shear varies in time, the growth of large perturbations is prevented in all directions, and rupture can hence be impeded.
\end{abstract}
\maketitle

\section{\label{sec:level1}Introduction}
Thin liquid films are found in many biological systems such as the human tear film \cite{WongFattRadke1996,SujaMossigeFuller2022} as well as in modern micro- and nanofabrication processes such as multilayer coextrusion of polymers \cite{LiBaer2020,BironeauSollogoub2017}. The stability of such films is often an important consideration, with hole formation due to film rupture often being undesirable \cite{BironeauSollogoub2017,craster2009}. In other applications, the rupture and dewetting of thin liquid films is intended, and can be manipulated in order to fabricate patterned materials \cite{FerrellHansford2007,DharaMukherjee2018}. 

It has been known for over half a century that the rupture of a thin liquid film can be caused by the amplification of small interfacial perturbations by long-range intermolecular van der Waals forces. In early works by Vrij \cite{Vrij1966} and Sheludko \cite{SHELUDKO1967}, the attractive surface interaction due to the van der Waals forces was represented by a thickness-dependent potential, and thermodynamic approaches were used to derive a critical wavelength above which perturbations to the flat film profile are unstable. This potential was later incorporated in a hydrodynamic model, which enabled derivation of the growth rate of a surface perturbation as a function of its wavelength, thus allowing the estimation of the rupture time \cite{RuckensteinJain1974}. 

The hydrodynamic model has been simplified by using the lubrication approximation to derive the so-called thin film equation, which describes the spatiotemporal evolution of the liquid film height with a single, highly nonlinear partial differential equation \cite{WilliamsDavis1982,OronBankoff1997}. This has led to numerous analytical and numerical works which have shown that the growth of surface perturbations can be separated into two regimes: an initial linear regime during which the film height has not yet deviated significantly from its initial value, and a subsequent nonlinear regime during which the growth of perturbations is greatly accelerated \cite{WilliamsDavis1982,SharmaRuck1986}. Zhang and Lister \cite{ZhangLister1999} showed that the nonlinear late-stage dynamics of rupture due to a disjoining pressure are governed by a similarity solution, with the minimum film thickness rapidly decreasing according to a power law. Deviation from an idealized disjoining pressure derived from the Lennard-Jones potential has been shown to produce discretely self-similar solutions  \cite{DallastonKalliadasis2018}. Such a process was also studied for analogous elastic interfaces \cite{carlson2015physfluids}. Beyond the initiation of film rupture, thin film models have also been used to  describe the droplet patterns formed during dewetting \cite{BeckerJacobsMecke2003}.

Van der Waals forces leading to liquid film rupture on solid surfaces only become significant when the film thickness is on the order of tens of nanometers~\cite{JacobsSeemannHerminghaus2008}. At these length scales and for common fluids at ambient temperatures, thermally driven molecular motion may cause significant fluctuations of the film height. Thus, it is natural to consider what role these microscopic thermal fluctuations play during the rupture process. In order to take these into account, a stochastic version of the thin-film equation was derived from the Navier-Stokes equation with an additional random stress tensor \cite{davidovitch2005,grun2006}. This formulation has been used to describe other thin film processes such as the spreading of a viscous bump under an elastic plate \cite{davidovitch2005,carlson_2018,pedersen_niven_salez_dalnoki-veress_carlson_2019} and the transport of solutes through nanopores \cite{marbach_dean_bocquet_2018}. Numerical and theoretical studies using the stochastic thin film equation in two dimensions have shown that random thermal fluctuations generate perturbations to an initially flat film profile, which eventually coarsen and approach the wavelength of maximum growth obtained from the linear stability analysis of the fluctuation-free system \cite{DiezFernandez2016,NesicKondic2015,MeckeRauscher_2005}. Simulations have also shown that increasing the fluctuation intensity decreases the rupture time and creates a more nonuniform pattern of droplet sizes after dewetting \cite{duran-olivenciakalliadasis,NesicKondic2015,ShahKreutzer}. Nevertheless, these thermal effects seem important only at the early stage of the  rupture process, with the late-stage behaviour being unaffected due to the dominance of the van der Waals force \cite{Zhaosprittles2022}.

One way to modify the dynamics of a thin liquid film is by applying a shear flow to it -- a situation widely encountered in industrial applications. Interestingly, shear is expected to dampen the amplitude of thermal interface fluctuations \cite{Derks2006,ThiebaudBickel2015,Bresson2017}, which might in turn influence the rupture time, \textit{i.e.}, the time it takes for the interface to touch down on the substrate. Numerical studies of deterministic thin film rupture in two dimensions have indeed shown that the presence of unidirectional shear can delay rupture when the interface has unstable perturbations \cite{KalpathyKumar2010,davis_gratton_davis_2010,Kadri2021}. To the authors' knowledge, experiments have not yet demonstrated the theorized rupture-suppressing effect of shear, but have shown that shear changes the morphology of the holes created when a polymer film dewets \cite{DmochowskaMiquelard-Garnier2022}. 

In this article, we solve numerically the stochastic thin film equation to improve our understanding of how shear affects the rupture dynamics of nanometric liquid films. By studying the film rupture for various combinations of fluctuation intensity, shear rate, and film thickness, we delineate the mechanisms by which rupture is affected in both two dimensions (2D) and three dimensions (3D). 

\section{Mathematical modeling and numerical methods} \label{sec:model}
\begin{figure*}
\includegraphics[width=\textwidth]{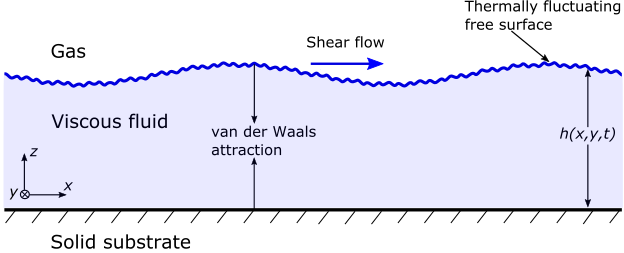}
\caption{\label{fig:schematic}Schematic of the studied physical system. A thin layer of viscous fluid with thickness $h(x,y,t)$ rests upon a flat solid substrate. A positive disjoining pressure, $\Pi (x,y,t)$, attracts the free surface to the solid substrate. The interface dynamics are also affected by shear and thermal fluctuations.}
\end{figure*}

\subsection{\label{sec:Equations}Stochastic thin film equation}

We consider the dynamics of a thin viscous liquid film as shown in Fig. \eqref{fig:schematic}. The dynamics of the film is represented by the spatiotemporal evolution of its height $h(x,y,t)$. For an initially flat film with height $h_0$, we define the perturbation to the film height as $\delta h=h(x,y,t)-h_0$. At any point in time, the minimum height of the film is denoted $h_{min}.$ Since we are interested in films where the height $h$ is much smaller than the horizontal length scale, the viscous flow profile in the fluid layer can be described by the lubrication approximation \cite{batchelor_2000}. A flux balance for a slice of fluid in the film then gives us the thin film equation describing how $h(x,y,t)$ varies with the pressure profile $p(x,y,t)$ \cite{OronBankoff1997}. In our system, the pressure $p(x,y,t)$ has contributions arising from the Laplace pressure due to the curvature of the free surface, the disjoining pressure $A^*/[6 \pi h^3(x,y,t)]$ resulting from the van der Waals interactions \cite{ISRAELACHVILI2011}, where $A^*$ is the Hamaker constant, and an additional stochastic stress arising from the thermal fluctuations in the fluid. The shear force is implemented in the tangential stress boundary condition at the film surface, taking the form $\mu\partial \textbf{u}/\partial z\rvert _{z=h(x,y,t)} = \boldsymbol{\tau}$, where $\textbf{u}(x,y,z,t)$ is the instantaneous velocity, $\mu$ is the dynamic viscosity, and $\boldsymbol{\tau}=\tau\boldsymbol{e}_\tau$ is the specified shear stress at the interface, with magnitude $\tau$ and oriented along the unit vector $\boldsymbol{e}_\tau$. Assuming incompressible flow of a Newtonian fluid, small Reynolds number and small slope of the film profile, $\nabla h(x,y) \ll 1 $, the thin-film equation reads:
  \begin{eqnarray}
    0=\frac{\partial h(x,y,t)}{\partial t} +  \nabla \cdot \left[\underbrace{\frac{\gamma}{3 \mu}h^3(x,y,t) \nabla\nabla ^2h(x,y,t)}_\text{Laplace pressure} + \underbrace{\frac{A^*}{6\pi\mu h(x,y,t)}\nabla h(x,y,t)}_\text{Disjoining pressure}\right] \nonumber \\+ \underbrace{\frac{\tau}{\mu} \boldsymbol{e}_\tau\cdot h(x,y,t)\nabla h(x,y,t)}_\text{Shear force}  + \underbrace{\sqrt{\frac{k_B T}{6\mu }}\nabla \cdot\left[h^{3/2}(x,y,t)\boldsymbol{\eta}(x,y,t)\right]}_\text{Stochastic force},
\label{eq:dimTF}
\end{eqnarray}
where $\gamma$ is the surface tension coefficient, $k_B$ is Boltzmann's constant and $T$ is the temperature. The last term in the thin-film equation accounts for the thermal fluctuations in the fluid, where $\boldsymbol{\eta}(x,y,t)$ is a random vector in the $(x,y)$-plane, the two components, $\eta_i$ with $i=x,y$, of which being 
independent delta-correlated spatiotemporal Gaussian noises with null averages \cite{MeckeRauscher_2005,davidovitch2005}, \textit{i.e.} $\langle\eta_i(x,y,t)\rangle=0$ and $\langle\eta_i(x,y,t)\eta_j(x',y',t')\rangle=\delta_{ij}\delta(x-x')\delta(y-y')\delta(t-t')$, where $\delta_{ij}$ is the Kronecker symbol and $\delta$ is the Dirac distribution.

We nondimensionalize Eq. \eqref{eq:dimTF} by introducing the scaling relations:
\[h=\bar{h}\sqrt{\frac{A^*}{2\pi \gamma}}, \hspace{10pt} x=\bar{x}\sqrt{\frac{A^*}{2\pi \gamma}},
\hspace{10pt} y=\bar{y}\sqrt{\frac{A^*}{2\pi \gamma}},
\hspace{10pt}
t=\bar{t}\sqrt{\frac{9A^*\mu^2}{2\pi \gamma^3}},\]

\[\boldsymbol{\eta}=\bar{\boldsymbol{\eta}}\left(\frac{8\pi^3}{9}\frac{\gamma^5}{A^{*3}\mu^2}\right)^{1/4}
, \hspace{10pt} \boldsymbol{\tau}=\bar{\tau}\boldsymbol{{e_{\tau}}}\sqrt{\frac{2\pi \gamma^3}{9A^*}}\]
where the bars indicate dimensionless quantities. Both vertical and horizontal lengths are nondimensionalized by a physical length scale, $\sqrt{A^*/(2\pi \gamma)}$, representing the characteristic film thickness at which the disjoining and capillary effects balance. We are mainly interested in the behavior of films with dimensionless initial thickness, $\bar{h}_0 \approx 1$. Films with $\bar{h}_0 \gg 1$ will not be significantly affected by the long-range van der Waals forces, while films with $\bar{h}_0 \ll 1$ will rupture almost instantaneously. The time is normalized by the aforementioned length scale divided by the capillary velocity $\gamma/(3\mu)$. The dimensionless stochastic thin film equation thus reads:

\begin{eqnarray} \label{eq:TFQB}
    0=\frac{\partial h(x,y,t)}{\partial t} +  \nabla\cdot\left[h^3(x,y,t)\nabla\nabla ^2h(x,t) + \frac{1}{h(x,t)}\nabla h(x,y,t)\right] \nonumber \\+B\boldsymbol{e}_\tau\cdot h(x,y,t)\nabla h(x,y,t) + Q  \nabla\cdot\left[h^{3/2}(x,y,t)\boldsymbol{\eta}(x,y,t)\right],
\end{eqnarray}
 where we have dropped the bars for simplicity, and introduced two dimensionless numbers: 
\begin{equation}
B=\bar{\tau}=\tau \sqrt{\frac{9A^*}{2 \pi \gamma ^3}},
\end{equation}
\begin{equation}
Q=\sqrt{\frac{\pi k_B T}{A^*}},
\end{equation}
representing the dimensionless shear force, and the dimensionless thermal roughness (or the ratio between thermal and disjoining energies), respectively. Along with the initial film height $h_0$, these parameters  define the thin-film dynamics.

We note that, when we model thin films in 2D (\textit{i.e.} when there is invariance in the $y$-direction), the fluctuation vector $\boldsymbol{\eta}$ needs to be adapted due to the inherently three-dimensional nature of thermal fluctuations. Eq.~(\ref{eq:TFQB}) remains valid, but $\boldsymbol{\eta}(x,t)$ is now a random vector in the $x$-direction only, the single  component $\eta$ of which being spatiotemporal Gaussian noise with $\langle\eta(x,t)\rangle=0$ and $\langle\eta(x,t)\eta(x',t')\rangle=\delta(x-x')\delta(t-t')$. As a consequence, $Q$ is modified in 2D, as:
\begin{equation}
Q_{2D}=\sqrt{\frac{\pi k_B T}{\bar{w}A^*}},
\end{equation}
where $\bar{w}=w\sqrt{\frac{2\pi \gamma}{A^*}}$, with $w$ a new length scale representing the width in the y-direction. 

\subsection{\label{sec:Numerics}Finite element solver}

We solve the dimensionless stochastic thin film equation (Eq. \eqref{eq:TFQB}) using the finite element method. In all simulations, the initial condition is a  flat film with dimensionless thickness $h_0$. Both in 2D and 3D, periodic boundary conditions are imposed in the horizontal directions. The domain size in the horizontal directions is always much greater than the thickness of the film to ensure the validity of the lubrication approximation.

To solve Eq.~(2), the order of the partial differential equation is first reduced by introducing the film curvature $\nabla ^2 h(x,y,t)$ as a separate variable. Eq. \eqref{eq:TFQB} then simplifies to a system of two coupled second order partial differential equations. These are then expressed in a weak  form where boundary terms disappear due to the periodic boundary conditions. The scalar fields $\nabla ^2 h(x,y,t)$ and $h(x,y,t)$ are discretized with linear elements and solved using a Newton solver from the FEniCS library \cite{LoggMardalEtAl2012}. For the 2D case, the simulations are realized on the domain $x=[0,65]$ with an equidistant grid spacing $\Delta x = 0.01$. In the 3D case, the domain is $x\times y=[0,64]\times [0,64]$, and the grid spacings are $\Delta x = \Delta y = 0.16$ for simulations used to generate the contour plots. For the data shown in other figures where many iterations are averaged, the grid spacing is increased to $\Delta x, \Delta y = 0.64$ in order to shorten the simulation time, and we have checked that the results are insensitive to the change in spatial resolution. Due to the fact that Eq.\eqref{eq:TFQB} involves a travelling wave, a second-order Crank-Nicholson scheme is required for numerical integration in order to prevent numerical dispersion for large values of $B$. The  size of the time step for all simulations shown is $\Delta t = 0.003$, with the exception of certain simulations involving $B>30$ as well as the adaptive time step simulations described later in this section and in appendix \ref{sec:adaptiv}. 

The stochastic term $\boldsymbol{\eta}(x,y,t)$ is implemented in python by assigning random numbers using the ``normal" function in the ``random" class of  NUMPY \cite{OliphantNumpy2006}. The values are drawn from a Gaussian distribution with zero mean and a variance of $1/(\Delta x \Delta t)$ in 2D and $1/(\Delta x^2 \Delta t)$ in 3D. At every time step, each component of $\boldsymbol{\eta}(x,y,t)$ is assigned a new value at every point in the mesh. Due to the stochastic nature of $\boldsymbol{\eta}(x,y,t)$, the film dynamics and rupture time vary somewhat between individual numerical simulations, despite all the input parameters being identical. In order to obtain statistically robust results for the wavelength and rupture time, we repeat the simulations and average the results $N=10$ times. Nevertheless, the variance of the rupture time and wavelength is generally so small that the trends we describe can be observed even for a single run.

For the non-stochastic simulations of the deterministic version of Eq. \ref{eq:TFQB}, an initial sinusoidal perturbation of amplitude $\delta h_0=0.001$ is imposed. In order to maintain the perdiodic boundary condition, the extent of the domain is set to 8 times the wavelength of the perturbation.

Previous works on shear-free films as well as the initial simulations of our system indicate that the highly nonlinear van der Waals force is completely dominant during the final stage of rupture when the minimum film height $h_{min}(t) = min(h(x,y,t))$ approaches zero \cite{ZhangLister1999, Zhaosprittles2022}. Due to the accelerated dynamics in this regime, a much smaller time step is required in order to capture the final moments before film rupture. To address this, we perform simulations with an adaptive time step, the results of which are shown in appendix \ref{sec:adaptiv}. In those simulations, the time step starts at $\Delta t = 0.1$  but is gradually reduced as rupture accelerates, reaching a minimum value of $\Delta t \approx 10^{-8} $. The grid spacing is also decreased to $\Delta x=0.001$ in order to resolve more of the details near the rupture point. As shown in appendix \ref{sec:adaptiv}, we recover the expected $h_{min}\sim (t_R-t)^{1/5}$ power law of Zhang and Lister \cite{ZhangLister1999}, and also observe that although this stage of rupture accounts for only a tiny fraction of the total duration of the film dynamics until rupture, it makes up most of the computational time due to the reduction in the time step. Our simulations with the adaptive timestep also show that the late stage before rupture seems to be independent of the thermal fluctuations and shear. We also note that close to rupture, lubrication theory cannot be used to describe the local flow \cite{morenoBozaMartinezCalvoSevilla2020}. Since we are mainly interested in understanding the instability growth rather than the details in the already characterized final instants before rupture, we decide to fix the time step and stop the simulation when $h_{min}$ reaches an arbitrary threshold $h^*=0.33$, which ensures that we capture the entirety of the early-stage dynamics without using unnecessary computational resources. We denote the time at which this occurs as the rupture time $t_R$.

\subsection{Linear stability analysis}
\label{subsec:teori}
 From the numerical solution of Eq. \eqref{eq:TFQB}, as will be described below, it is clear that the thermal fluctuations only play a major role in the evolution of $h_{min}$ during the first few time steps. As shown in appendix \ref{sec:adaptiv}, it is also clear that the late stage of pre-rupture film dynamics, when $h_0-h_{min}\gg 0.1$, is very short and as such does not make a significant contribution to the measured rupture time. Thus we presume that the rupture time of the film is primarily determined by how quickly perturbations grow while they are still too small for nonlinearities to dominate. Here, we perform a linear stability analysis on the deterministic version of Eq. \eqref{eq:TFQB} (\textit{i.e.} with $Q=0$) \cite{davis_gratton_davis_2010,Kadri2021}. We introduce a small sinusoidal perturbation to an initially flat film, of the form
\begin{equation}\label{eq:perturbation}
h(x,t)=h_0 + h'e^{ikx+\omega t}, 
\end{equation}
where $k$ is the dimensionless angular wave number, $\omega$ is the dimensionless complex growth rate, and $h'$ is the amplitude of the perturbation with $h_0\gg h'$.  Inserting Eq. \eqref{eq:perturbation} into the deterministic version of Eq. \eqref{eq:TFQB} and linearising the equation gives us the following dispersion relation
\begin{equation} \label{eq:growthrateandspeed}
\omega = \left({\frac{k^2}{h_0}-k^4 h_0^3}\right) - ik{Bh_0},
\end{equation}
with growth rate $\frac{k^2}{h_0}-k^4 h_0^3$ and wave speed $Bh_0$. From Eq. \eqref{eq:growthrateandspeed}, we expect that the effect of the shear within the linear regime is simply to cause a horizontal translation in the direction of shear with speed $Bh_0$. Moreover, only perturbations with a wavelength above a critical wavelength $\lambda _c = 2\pi h_0^2$ will grow in time, and the fastest growing mode has dimensionless wavelength
\begin{equation}\label{eq:lambdad}
\lambda _d = 2\pi \sqrt{2}h_0^2.
\end{equation}
The growth rate associated with this dominant wavelength is
\begin{equation}\label{eq:maxgrowth}
\omega _d = \frac{1}{4h_0^5}. 
\end{equation}
Presuming that the fluctuations trigger perturbations across the wavelength spectrum, we expect that $\lambda_d$ will be the dominant mode observed in our numerical results, and thus govern the change in minimum height of the film. If we only consider the minimum height in Eq. \eqref{eq:perturbation}, presume that the growth rate of this corresponds to Eq. \eqref{eq:maxgrowth}, and divide by $h_0$, we then get the following expression for how the minumum height of the film  varies with time:
\begin{equation}\label{eq:hminexpdev}
\frac{h_{min}(t)}{h_0}=1-\frac{h'}{h_0}e^{\omega _d t}.
\end{equation}
We consider rupture to occur at the time $t_R$ when $h_{min}$ reaches an arbitrary threshold value, $h^{*}=0.33$, at which the van der Waals forces become dominant:
\begin{equation}\label{eq:hstarexp}
\frac{h^{*}}{h_0}=1-\frac{h'}{h_0}e^{\omega _d t_R}.
\end{equation}
Solving this equation for $t_R$ and inserting our expression for the dominant growth rate in Eq. \eqref{eq:maxgrowth}, we then obtain the following estimate for the film rupture time 

\begin{equation} \label{eq:tRpred}
t_R=4h_0^5\ln\left(\frac{h_0-h^*}{h'}\right).
\end{equation}

\section{Results and Discussion}
\label{sec:resultater}

\subsection{Two-dimensional case}\label{subsec:2Dresults}

\begin{figure}
	 \centering
	\begin{tikzpicture}
   	\draw (0, 0) node[inner sep=0] (fig) {\includegraphics[width=0.47\textwidth]{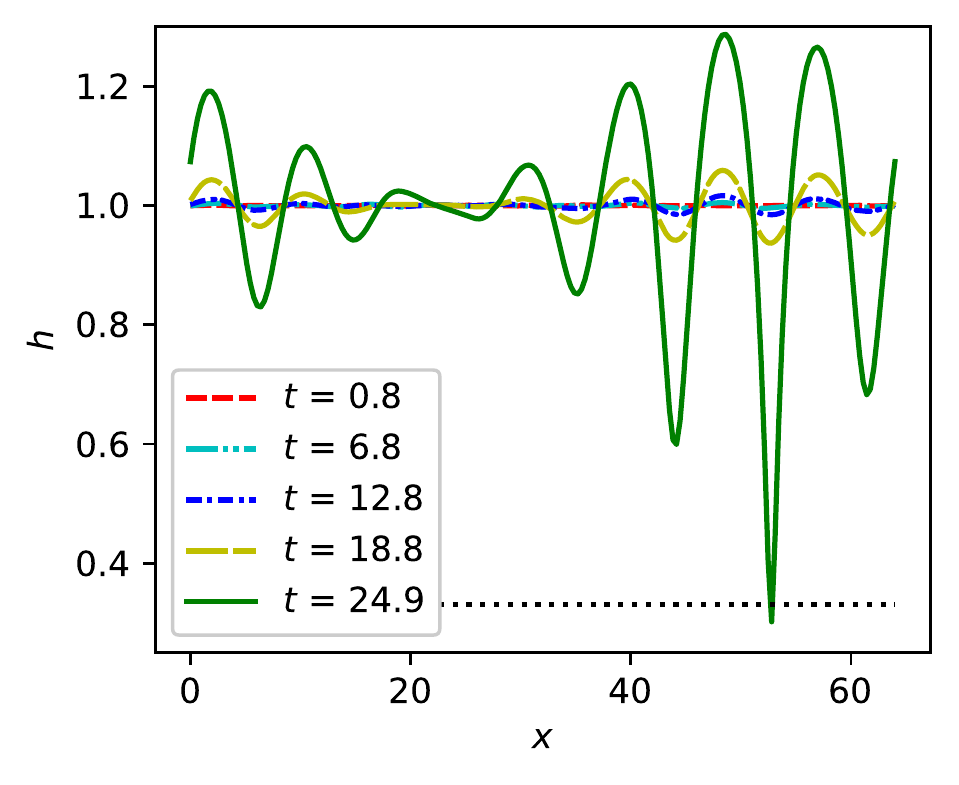}};
   	\node[] at (fig.north west){$(a)$};
 	\end{tikzpicture}
 	\begin{tikzpicture}
   	\draw (0, 0) node[inner sep=0] (fig) {\includegraphics[width=0.47\textwidth]{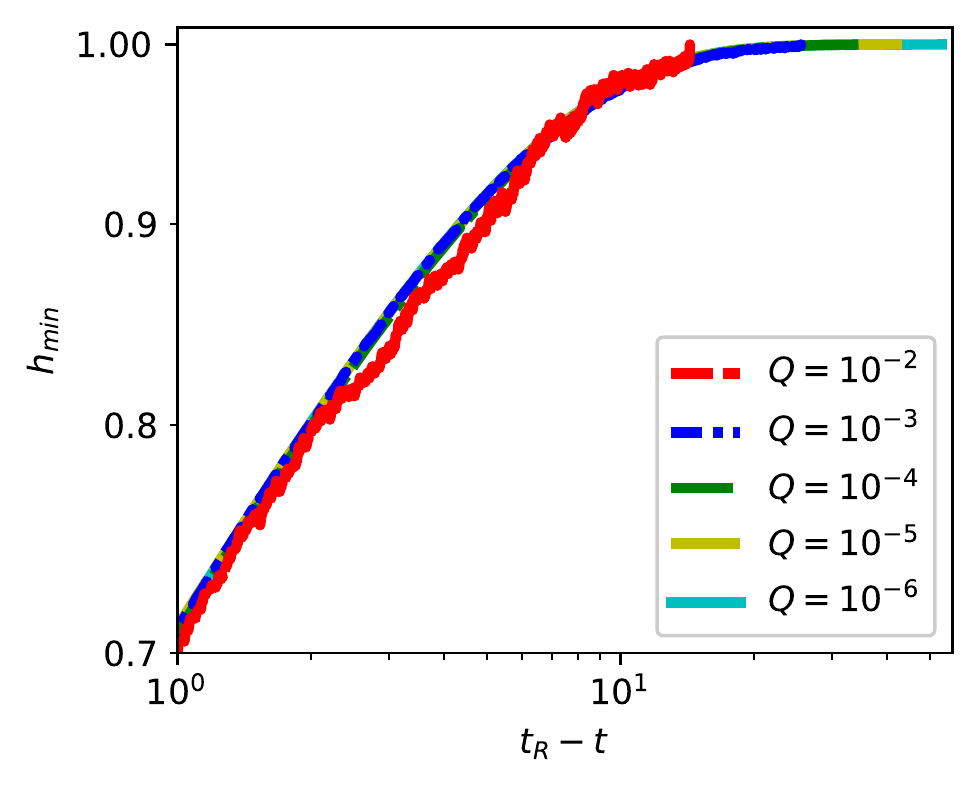}};
   	\node[] at (fig.north west){$(b)$};
 	\end{tikzpicture}

 	\begin{tikzpicture}
   	\draw (0, 0) node[inner sep=0] (fig) {\includegraphics[width=0.47\textwidth]{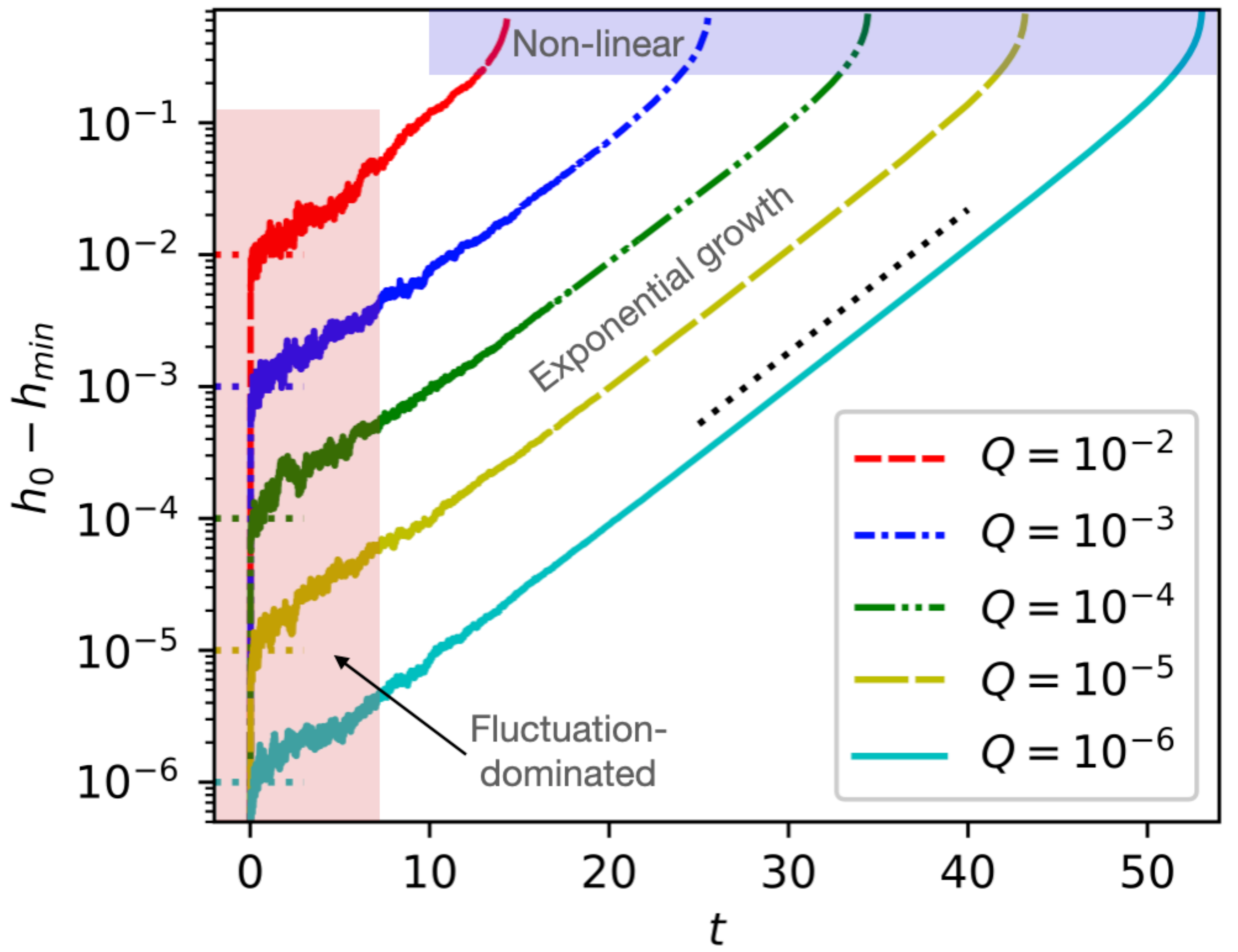}};
   	\node[] at (fig.north west){$(c)$};
 	\end{tikzpicture}
 	\begin{tikzpicture}
   	\draw (0, 0) node[inner sep=0] (fig) {\includegraphics[width=0.47\textwidth]{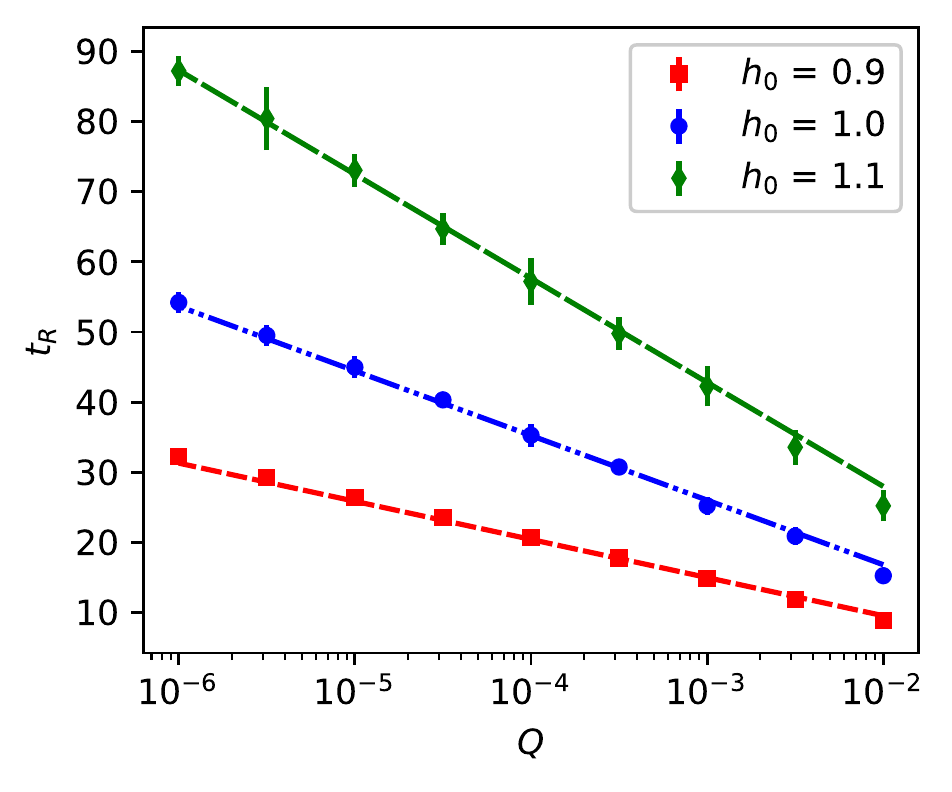}};
   	\node[] at (fig.north west){$(d)$};
 	\end{tikzpicture}

	\caption{$(a)$ Profiles of an unstable 2D thin film with $Q=10^{-3}$, $B=0$, and $h_0=1$ at various times. We stop the simulations when the minimum film height $h_{min}$ reaches an arbitrary threshold value $h^*=0.33$ (represented by the dotted black line), as described in section \ref{subsec:2Dresults}. $(b)$ Time evolution of the minimum film height, $h_{min}$, plotted for simulations with initial height $h_0=1.0$ and different fluctuation intensities, $Q$. $(c)$ Amplitude of the height perturbation as a function of time for the same simulations. The dotted black line has a slope corresponding to the prediction of Eq. \eqref{eq:hminexpdev} with arbitrary amplitude. The coloured dotted lines on the left indicate the value of $Q$ corresponding to each simulation. The three regimes mentioned in the text have been identified by eye as follows: the fluctuation-dominated regime at the begin of the simulations has been shaded red, while the area shaded in blue represents the nonlinear regime at the end of the simulation. $(d)$ Mean rupture time $t_R$ of a film with $B=0$ as a function of $Q$ for different values of $h_0$. The dashed lines represent Eq. \eqref{eq:tRpred_num} with no free parameter for each value of $h_0$. Error bars represent the standard deviation for a set of $N=20$ simulations for each data point. 
	\label{fig:2-D_noshear_multi}}
\end{figure}
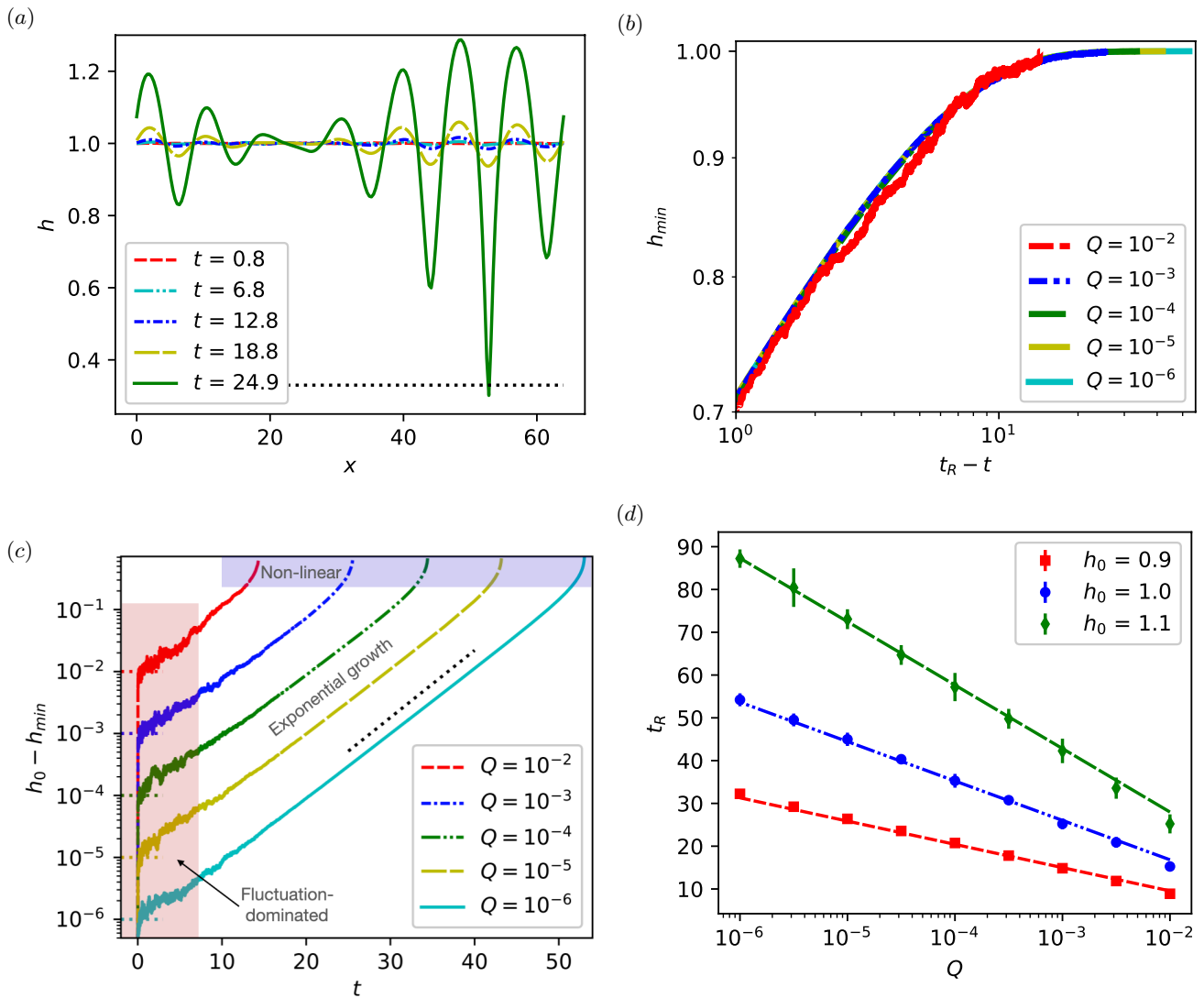

An example of the numerical solution of Eq. \eqref{eq:TFQB} without shear ($B=0$) in 2D is shown in Fig. \ref{fig:2-D_noshear_multi}$(a)$ in the form of height profiles along $x$ at different times. Thermal fluctuations eventually give rise to sinusoidal-like perturbations with a certain wavelength, which then grow until the film ruptures. For each film height $h_0$, the wavelength of the unstable film profile that develops closely matches the dominant wavelength $\lambda _d$ predicted in Eq. \eqref{eq:lambdad} and the rate of change of $h_{min}(t)$ closely matches the value predicted by Eq.\eqref{eq:maxgrowth}.

In Fig. \ref{fig:2-D_noshear_multi}(b-c), we can see how $h_{min}$ changes in time during individual simulation runs for different values of $Q$. As discussed above, $Q$ is proportional to the dimensionless thermal roughness, and we here present results as $Q$ varies from relatively small values until the limit where thermal fluctuations are so large that they rapidly cause rupture even without a disjoining pressure. Our results show that the primary effect of thermal fluctuations is to instigate a perturbation of a characteristic size $Q$ during the first time steps. Fluctuations  with approximately this size naturally continue to occur at all times, but after $t\approx 10$, the minimum height begins to decrease exponentially as one would expect in the deterministic case, and the fluctuations become insignificant in comparison to the growing sinusoidal perturbation. As shown in Fig. \ref{fig:2-D_noshear_multi}$(b)$, the time evolution of $h_{min}$ as the film approaches rupture is almost identical for different values of $Q$; the strength of the fluctuations seems to only determine how far back in time from the moment of rupture the curve starts at. In Fig. \ref{fig:2-D_noshear_multi}$(c)$ one can also see quite clearly that there are three distinct regimes during rupture: first a fluctuation-dominated phase that initiates a perturbation (shaded red in Fig. \ref{fig:2-D_noshear_multi}$(c)$), then a period of exponential growth of the roughly sinusoidal perturbation (unshaded in Fig. \ref{fig:2-D_noshear_multi}$(c)$), followed by a phase where the growth toward rupture is greatly accelerated due to the nonlinear effects of the disjoining pressure term in Eq.~\eqref{eq:TFQB} (shaded blue in Fig. \ref{fig:2-D_noshear_multi}$(c)$). The predominance of the exponential growth period during the rupture process seen in Fig. \ref{fig:2-D_noshear_multi}$(c)$ suggests that the rupture time of a film can be predicted by Eq. \eqref{eq:tRpred}. We also see from Fig. \ref{fig:2-D_noshear_multi}$(c)$ that the initial perturbation size, $h'$, is approximately equal to $Q$. Setting $h^*=0.33$, which appears to be a reasonable bound of the linear domain (\textit{i.e} with exponential growth of the instability), we can rewrite Eq. \eqref{eq:tRpred} as
\begin{equation} \label{eq:tRpred_num}
t_R=4h_0^5\ln\left(\frac{h_0-0.33}{Q}\right).
\end{equation}
 In \ref{fig:2-D_noshear_multi}$(d)$ we present the numerically measured rupture time from repeated simulations of our system as a function of $Q$ for various values of $h_0$. The predicted rupture time from Eq. \eqref{eq:tRpred_num}, represented by the straight dashed lines, provides an excellent fit to the data with no adjustable parameters.

\begin{figure}
	 \centering
   
	\begin{tikzpicture}
   
   	\draw (0, 0) node[inner sep=0] (fig) 
        {\includegraphics[width=0.47\textwidth]{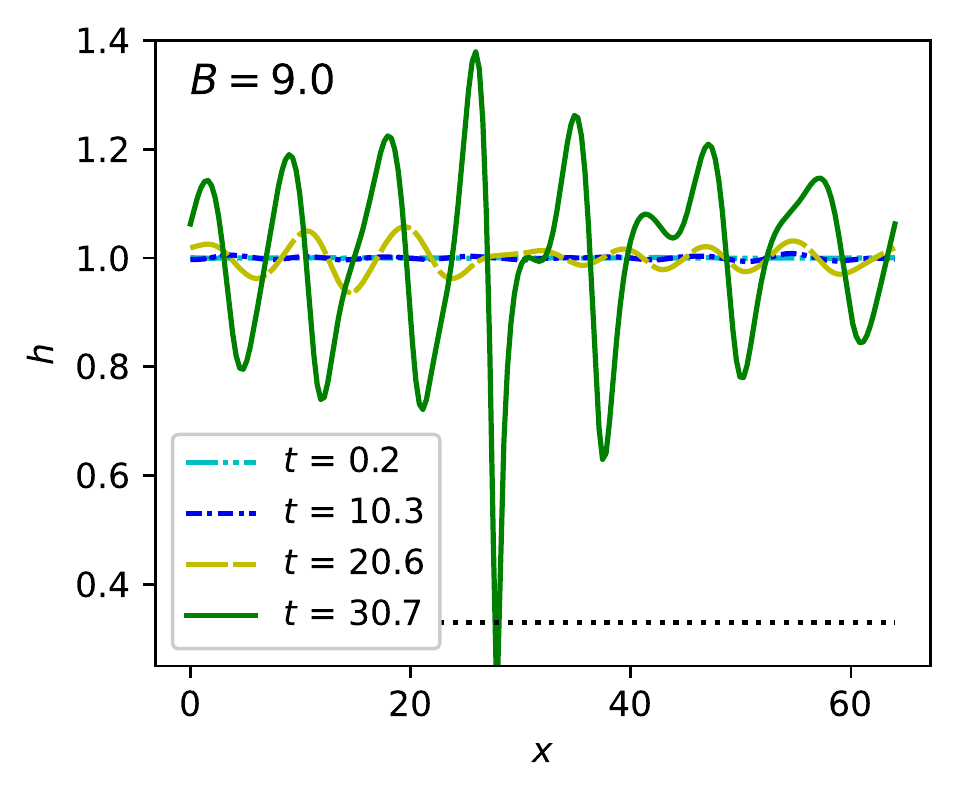}};
   	\node[] at (fig.north west){$(a)$};
 	\end{tikzpicture}
 	\begin{tikzpicture}
   	\draw (0, 0) node[inner sep=0] (fig) {\includegraphics[width=0.47\textwidth]{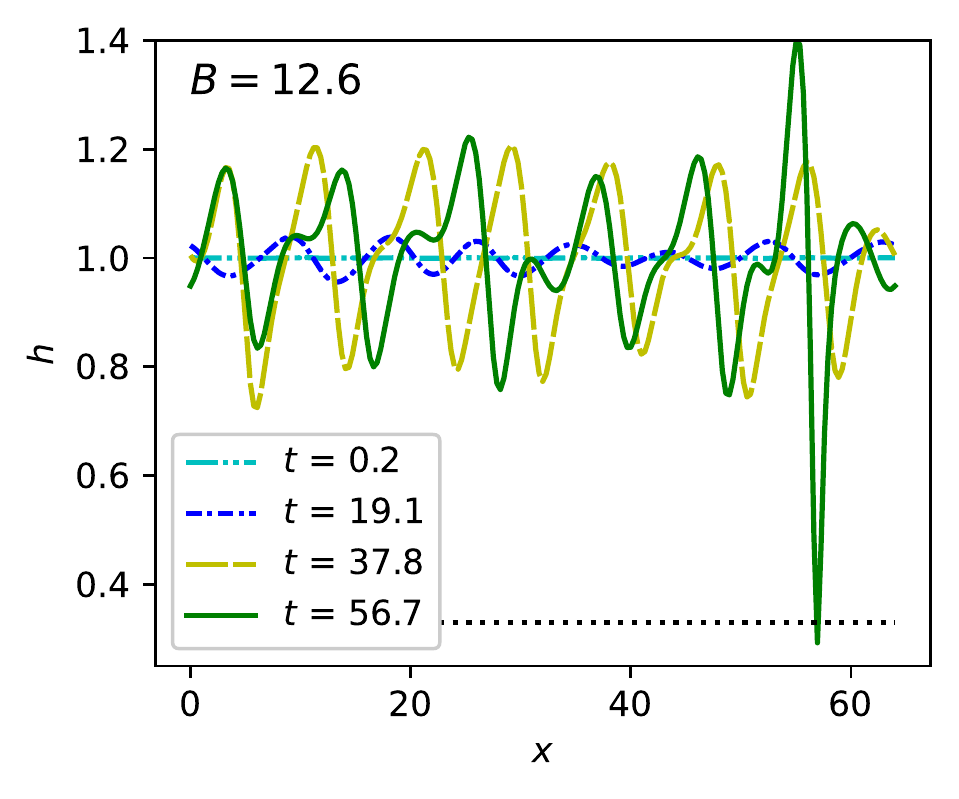}};
   	\node[] at (fig.north west){$(b)$};
 	\end{tikzpicture}

 	\begin{tikzpicture}
   	\draw (0, 0) node[inner sep=0] (fig) {\includegraphics[width=0.47\textwidth]{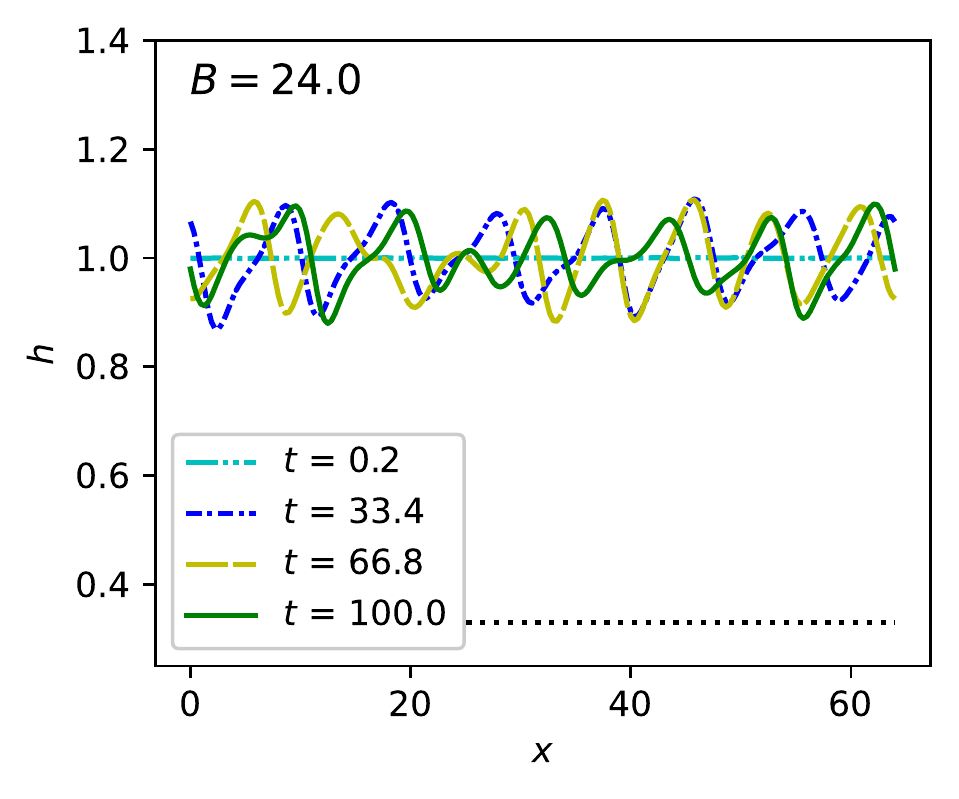}};
   	\node[] at (fig.north west){$(c)$};
 	\end{tikzpicture}
 	\begin{tikzpicture}
   	\draw (0, 0) node[inner sep=0] (fig) {\includegraphics[width=0.47\textwidth]{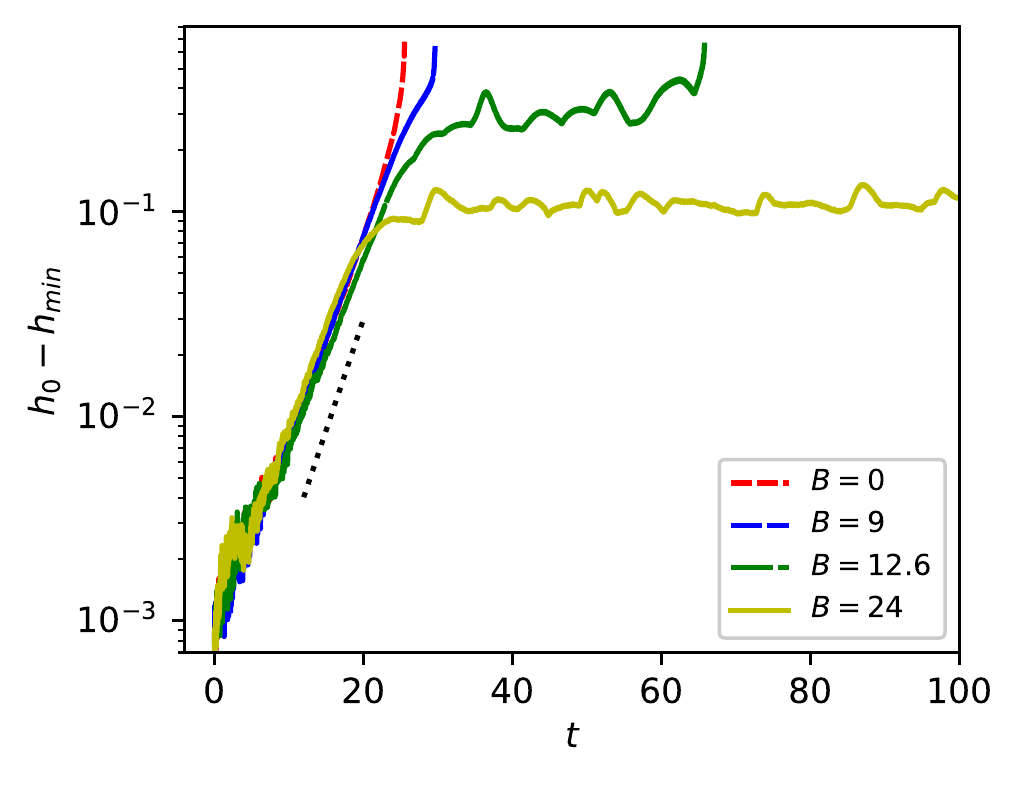}};
   	\node[] at (fig.north west){$(d)$};
 	\end{tikzpicture}

	\caption{$(a-c)$ Profiles of sheared 2D thin films with $Q=10^{-3}$ and $h_0=1$ at different times. The shear is from left to right in each case, with a non-dimensional size $B$ specified at the top of each figure. The dotted black line at the bottom indicates the threshold value $h^*=0.33$ at which the simulations are interrupted. $(a)$ $B=9.0$. $(b)$ $B=12.6$ $(c)$ $B=24.0$ $(d)$ Time evolution of the height perturbation $h_0-h_{min}(t)$ for different values of dimensionless shear stress $B$. The dotted black line indicates an exponential behaviour as predicted by Eq. \eqref{eq:hminexpdev}.
	\label{fig:2-D_shearprofiles}}
\end{figure}
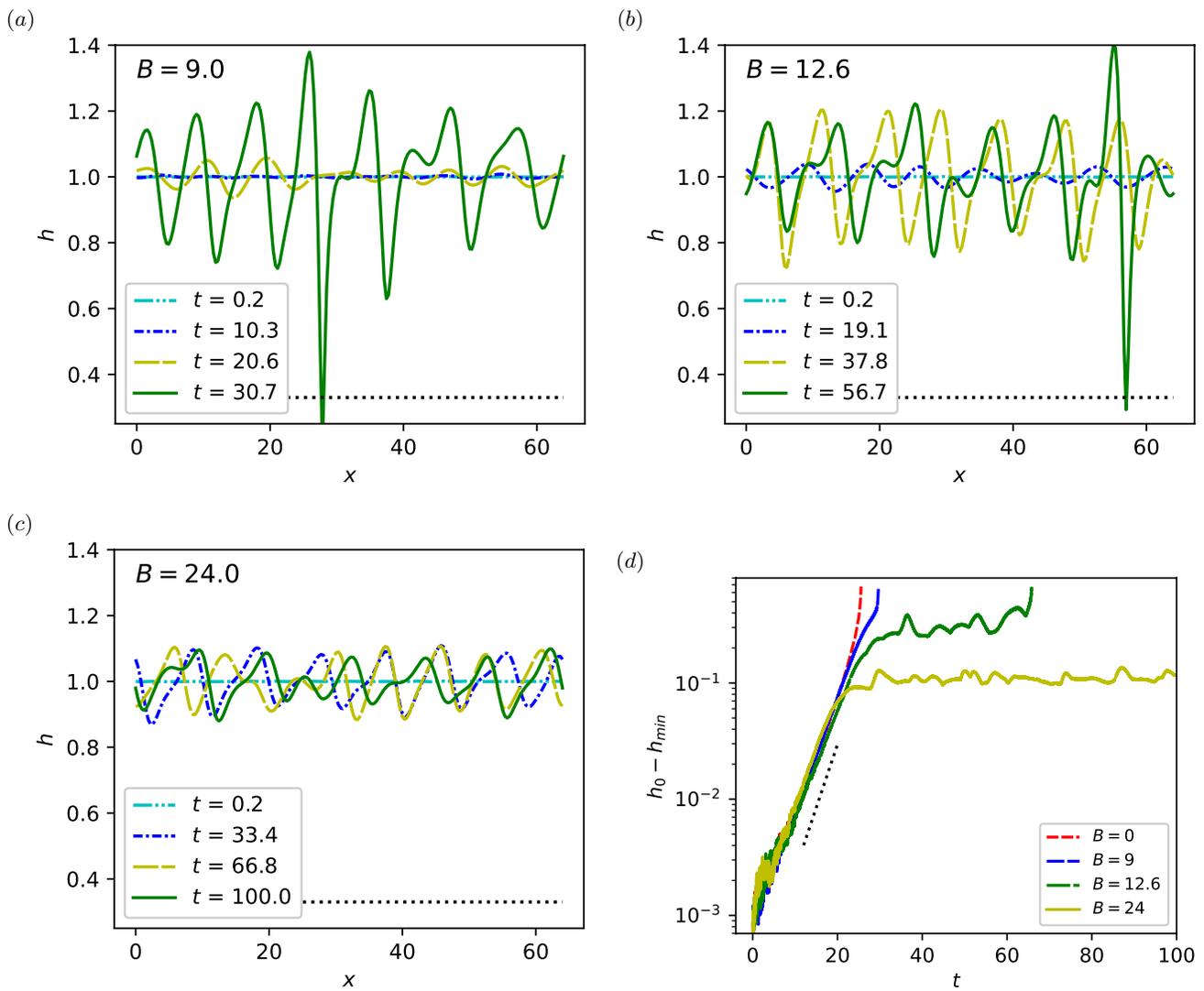

\begin{figure}
	 \centering
	\begin{tikzpicture}
   	\draw (0, 0) node[inner sep=0] (fig) {\includegraphics[width=0.47\textwidth]{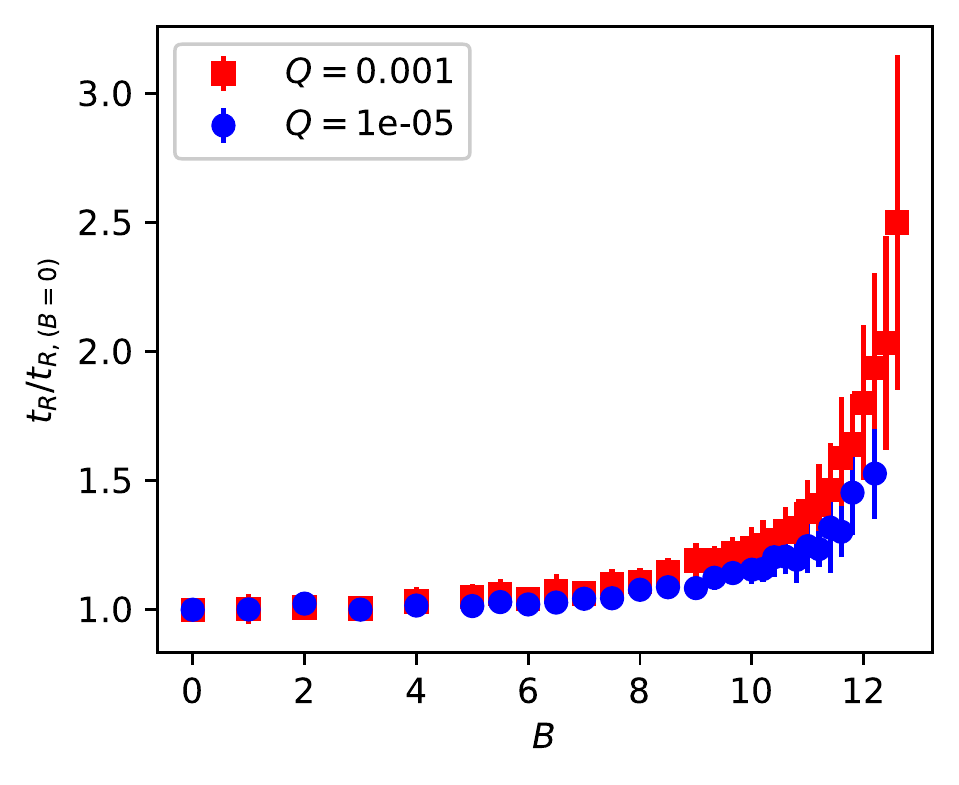}};
   	\node[] at (fig.north west){$(a)$};
 	\end{tikzpicture}
 	\begin{tikzpicture}
   	\draw (0, 0) node[inner sep=0] (fig) {\includegraphics[width=0.47\textwidth]{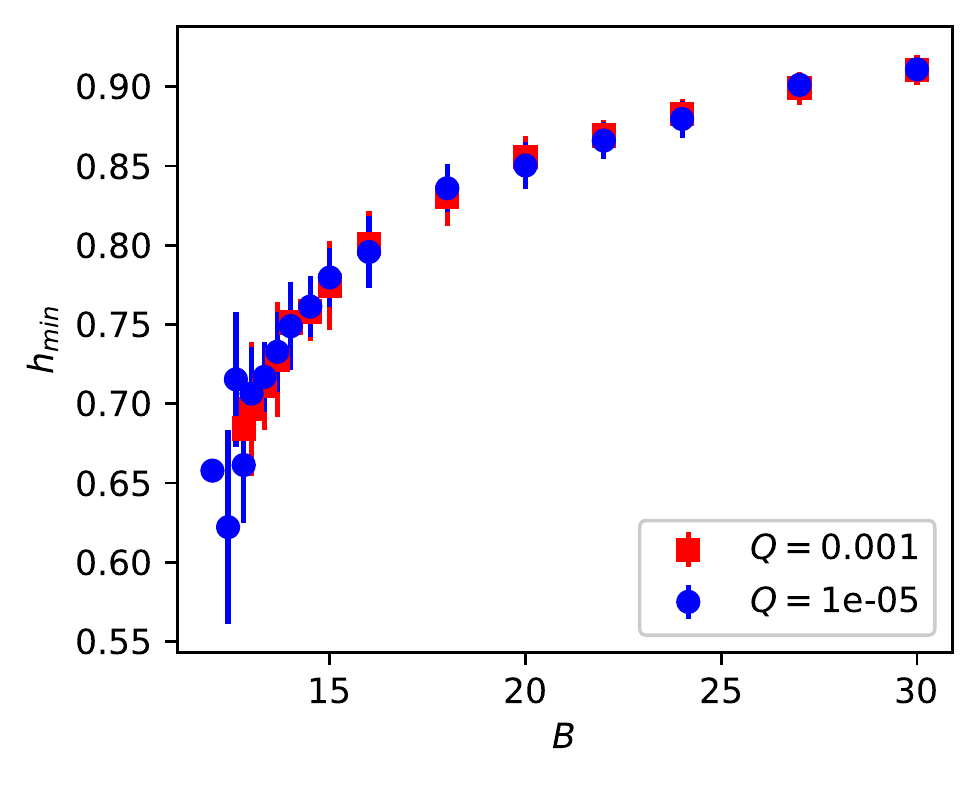}};
   	\node[] at (fig.north west){$(b)$};
 	\end{tikzpicture}

 	\begin{tikzpicture}
   	\draw (0, 0) node[inner sep=0] (fig) {\includegraphics[width=0.47\textwidth]{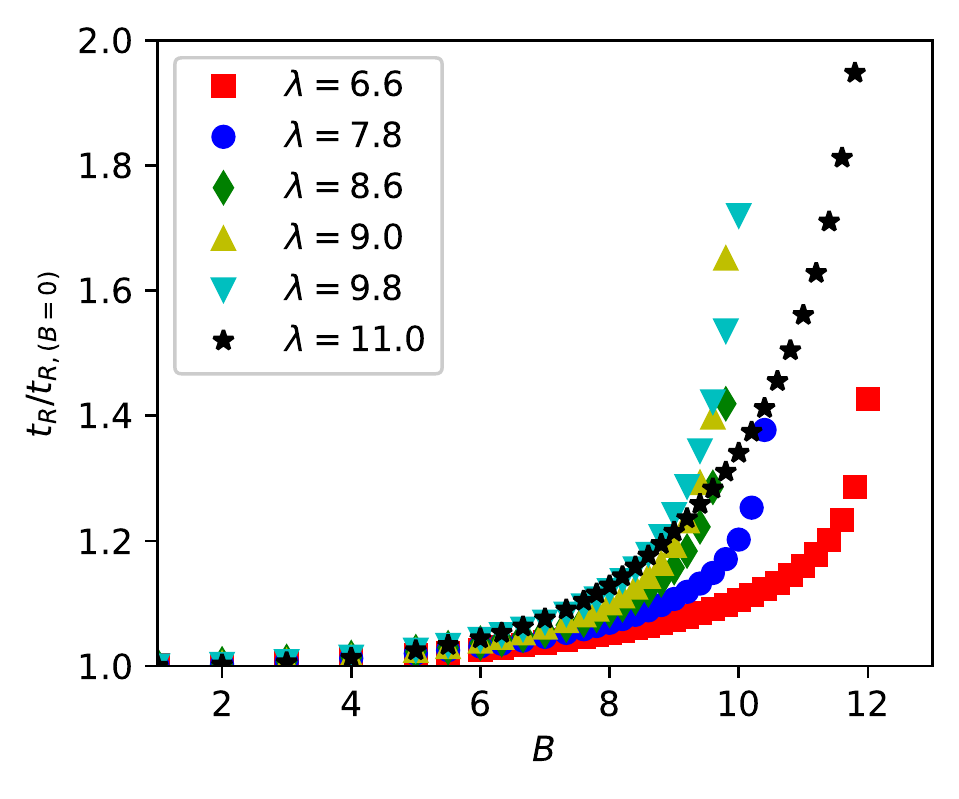}};
   	\node[] at (fig.north west){$(c)$};
 	\end{tikzpicture}
 	\begin{tikzpicture}
   	\draw (0, 0) node[inner sep=0] (fig) {\includegraphics[width=0.47\textwidth]{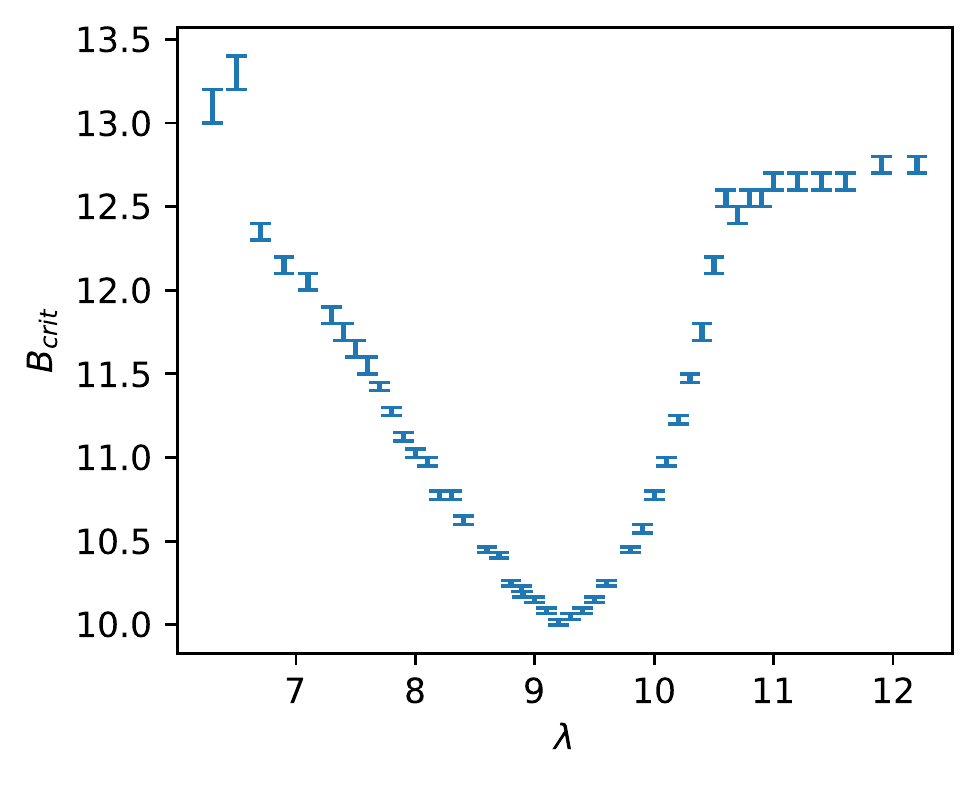}};
   	\node[] at (fig.north west){$(d)$};
 	\end{tikzpicture}

	\caption{$(a)$  Rupture time $t_R$ as a function of $B$ for two different values of $Q$. Rupture times are normalized by the rupture time for $B=0$. Error bars represent the standard deviation for a set of 20 simulations for each data point. $(b)$ Final value of the minimum film height as a function of $B$ for values above $B_{crit}$. Error bars represent the standard deviation for a set of 20 simulations for each data point.$(c)$ Rupture time as a function of $B$ for deterministic thin films with initial perturbations of varying wavelength $\lambda$. $(d)$ Plot of the critical shear required to suppress rupture within the simulated time period as a function of $\lambda$. The top and bottom of the error bars show the lowest shear that did not rupture and the largest shear that did rupture, respectively.  
	\label{fig:2-D_sheardata}}
\end{figure}

We now turn our attention to the effects of shear on the previous phenomenology. Eq. \eqref{eq:TFQB} is simulated with $B \neq 0$ and simulations are stopped either at rupture when $h \le 0.33$, or when the simulation has reached a maximum time of $t=100$. In Fig. \ref{fig:2-D_shearprofiles}$(a-c)$, we see snapshots of the film profile under shear with 3 different values of $B$. As predicted by the linear stability analysis (section \ref{subsec:teori}), the constant shear term leads to a simple horizontal translation of the perturbation in the shear direction, with speed $Bh_0$. We observe that the perturbations remain roughly sinusoidal-like during the exponential growth stage, and maintain a wavelength close to $\lambda_d$ as well as a growth rate of approximately $\omega _d$. In the nonlinear regime, however, they start to deform as the van der Waals forces become stronger, which gives rise to characteristic asymmetric shapes in the nonlinear stage. For $B \gtrsim 5$, the change in shape is noticeable, and is associated with a short period of decreased perturbation growth, which leads to a slight increase in rupture time, as shown in Figs. \ref{fig:2-D_shearprofiles}(a) and \ref{fig:2-D_shearprofiles}(d). As shear is increased to around $B \gtrsim 11$, however, the perturbations begin to stabilize when they reach a certain size, and a seemingly stable wave seems to form for a period of time, before it eventually ruptures, as can be seen in Figs. \ref{fig:2-D_shearprofiles}(b) and \ref{fig:2-D_shearprofiles}(d). For shear stronger than a critical value of $B \approx 12.5$, this stable translating perturbation does not rupture within the simulation time. We denote this critical value as $B_{crit}$. For $B > B_{crit}$, the perturbation waveform seemingly propagates indefinitely as shown in Figs. \ref{fig:2-D_shearprofiles}(c) and \ref{fig:2-D_shearprofiles}(d). 

In Fig. \ref{fig:2-D_shearprofiles}(d) it is apparent that imposing a shear below $B_{crit}$ can delay rupture to some extent. This effect is quantified in Fig. \ref{fig:2-D_sheardata}(a), where it can be seen that imposing $B > 5$ leads to a modest increase in rupture time, and that this effect is dramatically increased as $B$ approaches $B_{crit}$. Fig. \ref{fig:2-D_shearprofiles}(d) also suggests that beyond $B_{crit}$, the amplitude of the stable travelling wave generated under shear is dependent on the size of $B$. This is indeed the case, as is shown in Fig. \ref{fig:2-D_sheardata}(b). Just above $B_{crit}$, the minimum height of the final film profile is around $h_{min}=0.65$, but this increases as $B$ is increased, eventually converging towards a complete suppression of the perturbation as B approaches infinity.

Interestingly, we note that the observed value of $B_{crit} \approx 12.5$ is  larger than the value of 9.7 observed for the deterministic case with an initial perturbation of wavelength $\lambda_d$ by Davis et al. \cite{davis_gratton_davis_2010}. In Fig. \ref{fig:2-D_sheardata}(c) we plot the increase in rupture time with $B$ for deterministic simulations (i.e. Eq. \eqref{eq:TFQB} with $Q=0$ and a sinusoidal initial perturbation) for a few different perturbation wavelengths. In this plot we note that the critical shear for the dominant mode $\lambda_d$ is indeed around 9.7, but that larger and smaller wavelengths require more shear in order to suppress rupture. We propose that fluctuations enable rupture even beyond the critical shear of the dominant mode by triggering secondary modes once the dominant mode begins to stabilize. This is supported by the fact that the critical shear observed in the stochastic case approximately matches the largest values of $B_{crit}$ for large and small wavelengths, as is shown in Fig. \ref{fig:2-D_sheardata}(d)

In Fig. \ref{fig:2-D_shearprofiles}, we observe that the size of the initial decrease $h_0-h_{min}(0)$ in film height due to the stochastic term seems to be of size $Q$ regardless of the strength of the shear. This is rather unexpected~\cite{ThiebaudBickel2015} if the size of the initial perturbation is interpreted as the equilibrium thermal roughness. We note however, that the sinusoidal perturbations that lead to rupture form within the first few time steps of the simulation, long before the interface settles to an equilibrium roughness. This is discussed in more detail in appendix \ref{sec:shearroughness}, where we also show that the equilibrium thermal roughness is decreased by $B$ when there is no disjoining pressure.

\subsection{Three-dimensional case} \label{subsec:3Dresults}
\begin{figure}
	 \centering
\includegraphics[width=0.95\textwidth]{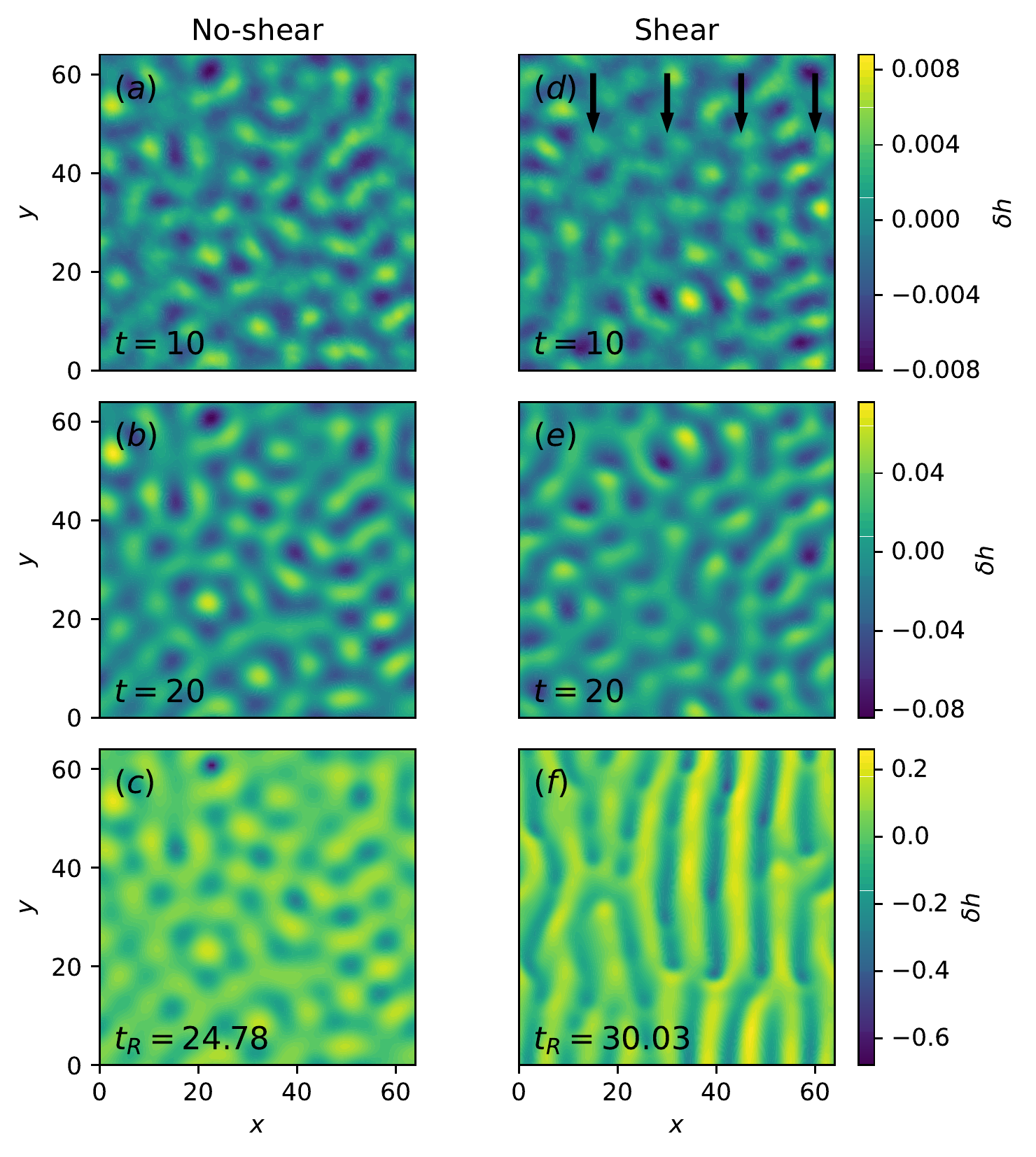}

	\caption{Contour plots illustrating the height $h(x,y,t)$ of 3D thin films for different times, with $Q=10^{-3}$. The color represents the deviation $\delta h(x,y) = h(x,y)-h_0$ from the initial film thickness $h_0$. $(a-c)$ No-shear ($B=0$) case, where $(c)$ shows the final time step at which the film ruptures. $(d-f)$ Shear ($B=30$, along $y$-axis) case, where (f) shows the final time step at which the film ruptures. Note that the direction of shear (as indicated by the black arrows in panel $(d)$) is not the same as in Fig. \ref{fig:schematic}.
	\label{fig:3Dprofiles}}
\end{figure}
Although our results in two dimensions shed light on 2D film rupture, it is not clear how these effects translate to physically realistic 3D films. To quantify the effect of spatial dimensions, we solve Eq. \eqref{eq:TFQB} to obtain 3D film profiles, obtaining the results shown in Fig. \ref{fig:3Dprofiles}. In Fig. \ref{fig:3Dprofiles}$(a-c)$, we note that the shear-free rupture process is both qualitatively and quantitatively similar to the observations in two dimensions (see Fig. \ref{fig:2-D_noshear_multi}$(a)$): the stochastic fluctuations give rise to small perturbations that coarsen until a characteristic size given by $\lambda_d$ and subsequently grow over time. The surface depressions that lead to rupture have a circular-like cross-section, but grow with the same growth rate as in the 2D case (Eq. \eqref{eq:maxgrowth}). The rupture time is still fairly well described by Eq. \eqref{eq:tRpred}.

\begin{figure}
	 \centering
	\begin{tikzpicture}
   	\draw (0, 0) node[inner sep=0] (fig) {\includegraphics[width=0.47\textwidth]{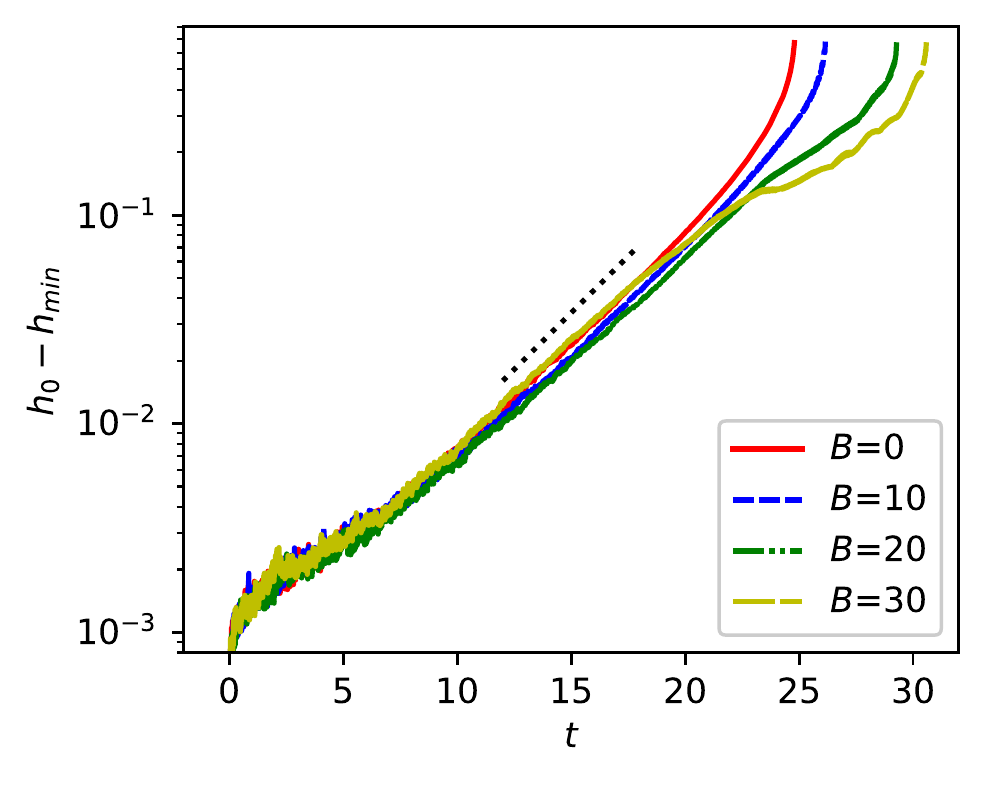}};
   	\node[] at (fig.north west){$(a)$};
 	\end{tikzpicture}
 	\begin{tikzpicture}
   	\draw (0, 0) node[inner sep=0] (fig) {\includegraphics[width=0.47\textwidth]{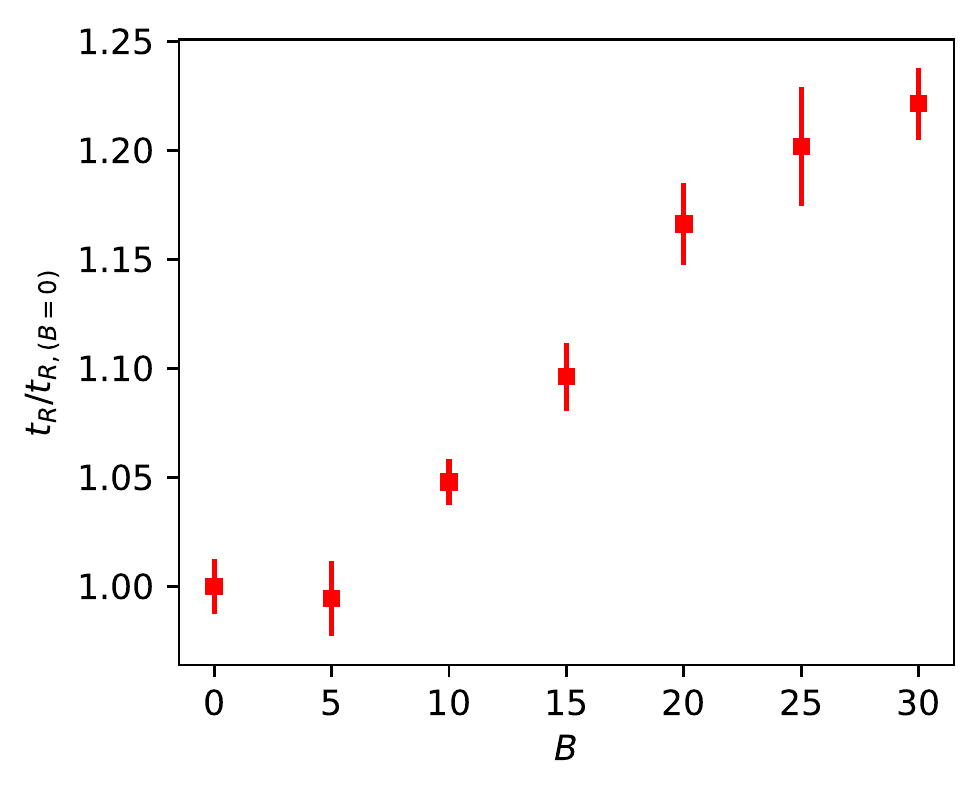}};
   	\node[] at (fig.north west){$(b)$};
 	\end{tikzpicture}

	\caption{$(a)$ Time evolution of the height perturbation $h_0-h_{min}(t)$ for different values of $B$ in 3D simulations, with unidirectional shear and for $Q=10^{-3}$,$h_0=1$. The dotted black line line indicates exponential behaviour with a growth rate given by Eq. \eqref{eq:maxgrowth}. $(b)$ The rupture time as a function of shear magnitude, where the error bars represent the standard deviation for a set of 5 simulations for each value of $B$.  
	\label{fig:3Dunidirectionalshear_dhvt&tRvB}}
\end{figure}

In panels $(d-f)$ of Fig. \ref{fig:3Dprofiles}, we can see how a strong unidirectional shear in the $x$-direction affects the three-dimensional film rupture process. In this case, the shear force has a suppressing effect on perturbations along the shear direction, but perturbations grow freely in the transverse direction. As the perturbations grow to a significant size, they begin to align with the direction of shear, and eventually form a clear pattern of ridges. These ridges then deepen as the film rupture process proceeds. In Fig. \ref{fig:3Dunidirectionalshear_dhvt&tRvB}$(a)$, we show how the minimum film height evolves in time for 3D simulations of rupture with unidirectional shear of varying strength. As in the 2D case, the growth rate in the exponential regime is close to $\omega_d$ predicted by the linear stability analysis. In contrast to the 2D case of Fig. \ref{fig:2-D_shearprofiles}, however, unidirectional shear does not prevent rupture from occurring in a 3D film. Nevertheless, for $B>B_{crit}$ there is a small increase in rupture time, as shown in Fig. \ref{fig:3Dunidirectionalshear_dhvt&tRvB}$(b)$, caused by the fact that the perturbations need to rearrange themselves when they are suppressed in the shear direction. 

\begin{figure}
	 \centering
\includegraphics[width=0.95\textwidth]{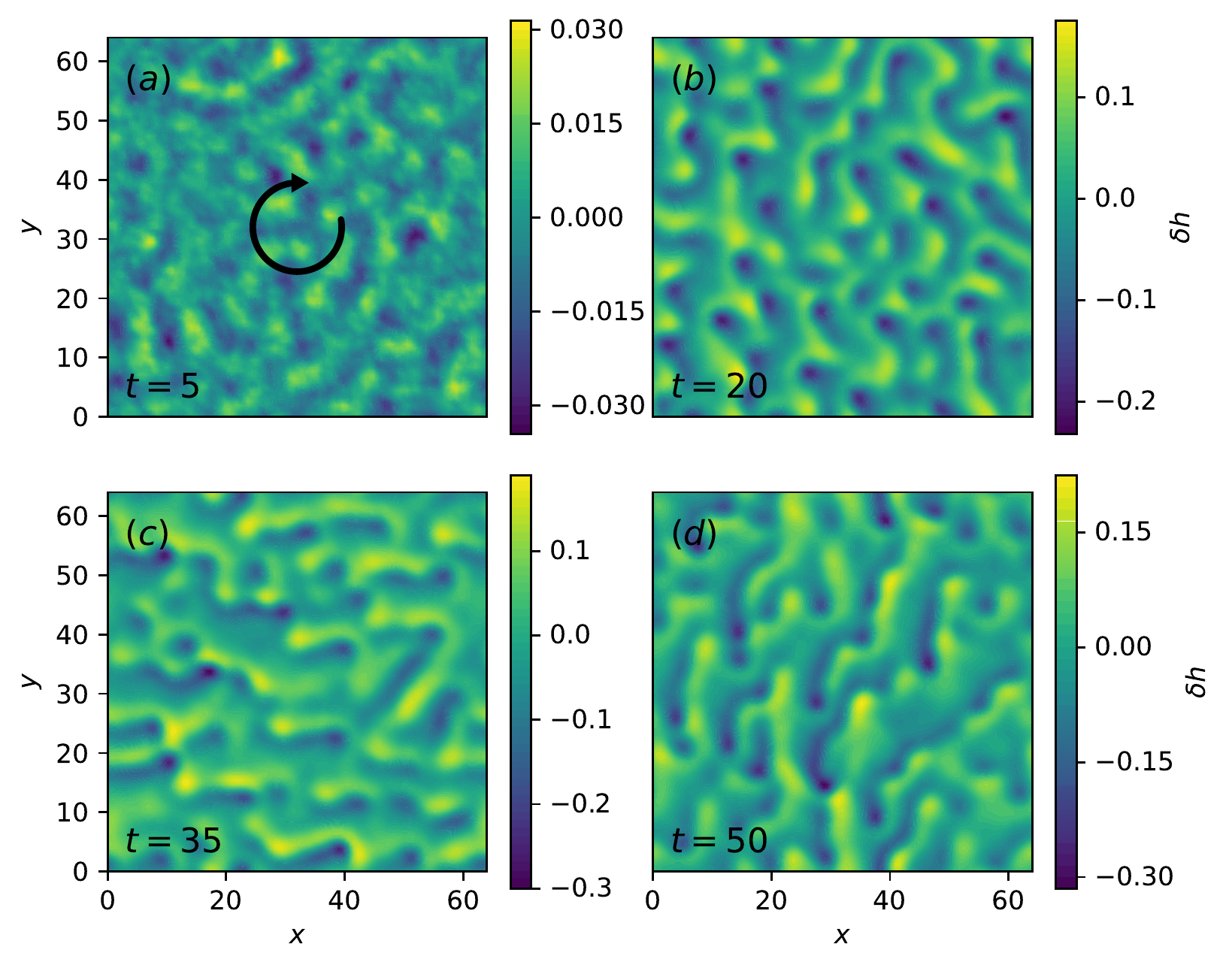}

	\caption{Contour plots illustrating the height $h(x,y,t)$ of a 3D thin film, subjected to a rotating shear force with strength $B=30$ and angular frequency $\Omega = 3\pi/32$ in the direction of the arrow in panel $(a)$, as a function of time, with $Q=10^{-2}$ and $h_0=1$. The color represents the deviation $\delta h(x,y) = h(x,y)-h_0$ from the initial film thickness $h_0$. 
	\label{fig:3Dprofiles_rot}}
\end{figure}

\begin{figure}[h]
	 \centering
	\begin{tikzpicture}
   	\draw (0, 0) node[inner sep=0] (fig) {\includegraphics[width=0.47\textwidth]{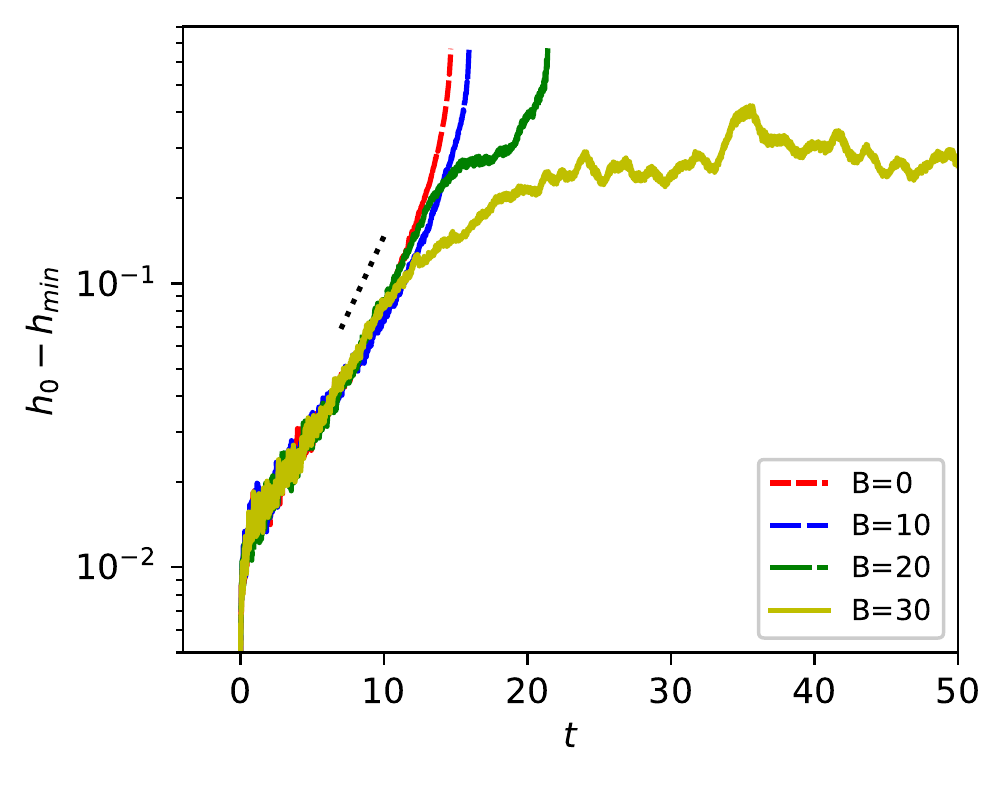}};
   	\node[] at (fig.north west){$(a)$};
 	\end{tikzpicture}
 	\begin{tikzpicture}
   	\draw (0, 0) node[inner sep=0] (fig) {\includegraphics[width=0.47\textwidth]{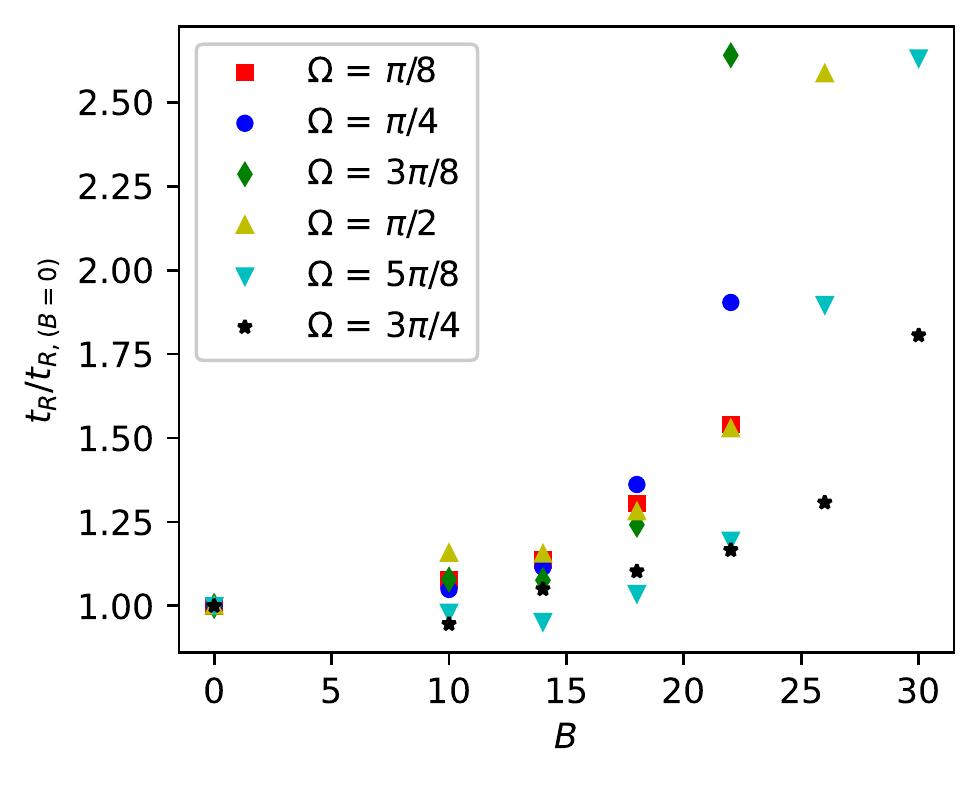}};
   	\node[] at (fig.north west){$(b)$};
 	\end{tikzpicture}

	\caption{$(a)$ Time evolution of the height perturbation $h_0-h_{min}(t)$ for different values of the shear parameter $B$ and angular frequency $\Omega=\pi/8$, in 3D simulations with rotating shear for $Q=10^{-2}$. The dotted black line line indicates exponential behaviour with a growth rate given by Eq. \eqref{eq:maxgrowth} and an arbitrary amplitude. $(b)$ Rupture time $t_R$ as a function of dimensionless shear stress $B$, for various angular frequencies $\Omega$.
	\label{fig:rotskjærdhvt&tRvB}}
\end{figure}

Since the film rupture time is weakly affected by unidirectional shear at the interface, we then attempted to impose a rotating shear stress, as was done previously by Davis \textit{et al.} \cite{davis_gratton_davis_2010}. To do so, Eq. \eqref{eq:TFQB} is again solved numerically, but the direction of the shear force, $\overrightarrow{e}_\tau$, is now varied sinusoidally in time with an angular frequency $\Omega$. The resulting rupture process can be seen in Fig. \ref{fig:3Dprofiles_rot}, and is further analyzed in Fig. \ref{fig:rotskjærdhvt&tRvB}. During the linear regime, the perturbations rotate as the direction of $\overrightarrow{e}_\tau$ changes, but grow much as they do without shear. Once the perturbations become big enough for nonlinear effects to emerge, the perturbations begin to be suppressed in the direction of shear, leading to a reorientation as in the unidirectional case. For very  small frequencies, $\Omega \ll 2\pi /t_{R,B=0}$, the change in direction is too slow to affect the rupture process significantly. When $\Omega$ is increased, however, the shear has time to switch directions and suppress the perturbations that have grown in the initially cross-shear direction. This leads to periodic reorientation of the perturbation profile with an angular frequency equal to $\Omega$, and an overall delay of rupture. For rapidly rotating shear, the perturbations remain essentially circular in cross-section and resemble the perturbations seen in the shear-free case shown in Fig. \ref{fig:3Dprofiles}$(a-c)$. If $B$ is sufficiently large, and $\Omega$ has an appropriate value, the perturbation can be completely suppressed until the maximum simulation time of 50, as is the case in Fig. \ref{fig:3Dprofiles_rot}.

For a given angular frequency, $\Omega$, the change in rupture time as $B$ is increased, as depicted in Fig. \ref{fig:rotskjærdhvt&tRvB}, is actually quite similar to the 2D case of section \ref{subsec:2Dresults}. For $B<10$, shear does not significantly delay rupture, but does so once $B\gtrapprox 15$ due to slower growth in the nonlinear regime. The critical value of $B$ required is significantly higher than that observed in 2D, and varies with $\Omega$. As seen in Fig. \ref{fig:rotskjærdhvt&tRvB}$(b)$, it seems that the rupture-delaying effect of shear is strongest when $\Omega \approx 3\pi/4$. For lower frequencies, the shear suppresses rupture in one direction, but does not change fast enough to suppress it in the other, whereas for high frequencies, the shear does not remain in one direction for long enough to suppress rupture in that direction.

\subsection{Implications for polymer processing} 
\label{subsec:Discussion}

Let us now discuss the implications of the numerical results obtained above in a practical context. We specifically consider a polymer processing method: nanolayer coextrusion. This innovative process is based on a series of layer multiplying elements (LME) which apply the baker's transformation (successive slicing and recombining) to a stratified polymer melt flow to achieve multilayer systems, made of up to thousands of alternating layers of two or more polymers, each having nanometric thicknesses \cite{Gholami20}. Many polymer pairs have been processed this way, whether glassy or semi-crystalline, miscible or immiscible and in most cases stable, continuous and regular layers with thicknesses below 50 nm have been obtained \cite{Rijal22,Lozay21,Nassar18}. Let us focus on the widely studied case of polystyrene (PS) / polymethylmetacrylate (PMMA) multilayers, for which it has been shown that optimized extrusion conditions can lead to continuous layers as thin as 20 nm \cite{BironeauSollogoub2017}.  Typical values for extrusion temperature, surface tension, dynamic viscosity and Hamaker constant can be set to 220°C, 1 mN/m \cite{Wu70}, 10000 Pa.s \cite{BironeauSollogoub2017} and 2.10$^{-18}$ J \cite{deSilva12,Kadri2021}, respectively. In this case, the thermal roughness at the interface has an amplitude  $\sqrt{k_B T/\gamma}\approx 2.5\cdot 10^{-9}$ m, and $Q\approx 10^{-1}$ which is above the values presented in this study. Nonetheless, from Figs. \ref{fig:2-D_noshear_multi} and \ref{fig:2-D_shearprofiles}, it can be hypothesized that though faster, rupture will occur similarly as with lower $Q$ values. Using Eq. \eqref{eq:tRpred_num} and putting back dimensions, the rupture time without shear for a layer having an initial thickness of 20 nm ($h_0 \approx 1.1$) would be in the 5 - 10 s range, hence much less than the typical processing time, between 1 to 2 minutes \cite{Bironeau2016}. The stabilizing effect of shear could explain the stability of nanometric layers (around 20 nm), as already discussed in \cite{Kadri2021}. Looking at the 2D case (Fig. \ref{fig:2-D_sheardata}(c)), a significant increase of the rupture time occurs when B is higher than the critical value around 12.5, corresponding to a shear rate in the order of 20 - 25 s$^{-1}$, that is easily reached in classical extrusion conditions \cite{Bironeau2016}. 
However, the novel 3D simulations of the present study in uniaxial shear suggest that the shear force, $B$, has in fact a limited influence on the rupture time, which appears to contradict the previous conclusions drawn in 2D. The following hypotheses can be made concerning processing, especially nanolayer coextrusion: the minimal thickness achieved after the LMEs is actually higher than the one discussed previously, nanometric thicknesses being only reached after the LMEs when the flow goes through the flat die. As the rupture time scales with the thickness to the power of 5, rupture may only occur when the polymer flow passes through this flat die, where the layer thickness reaches values below 100 nm. Interestingly, in the flat die, there is not only unidirectional shear but also a diverging radial flow, which may result in a more stabilizing situation as suggested by results presented in Figs. \ref{fig:3Dprofiles_rot} and \ref{fig:rotskjærdhvt&tRvB}. Besides, we stress that the effect of elongation has not been elucidated yet, and we shall also mention that the boundary conditions used here in a single layer differ quite substantially with those occurring in a multilayer flow \cite{Lenz07}.  

Finally, we comment on the 3D surface profiles showing the appearance of ridge-like and valley-like features aligned along the shear direction prior to rupture.
This pattern formation is reminiscent of the experimental observations by Dmochowska \textit{et al.} \cite{DmochowskaMiquelard-Garnier2022}, where a PS thin film sandwiched between two PMMA thicker layers was dewetted under shear. 
Though this study focused on the dewetting kinetics, \textit{i.e.} the hole growth after hole formation, it was observed that the dewetting holes take ellipsoidal shapes aligned with the local shear directions, contrary to the no-shear case \cite{ZHUMiquelardGarnier2016,Chebil18}.
The hole growth was accelerated in the shear direction while it remained similar to the no-shear case (with a slight increase caused by shear-thinning of the PMMA matrix) in the perpendicular direction. 

\section{Conclusion}
\label{sec:conclusion}

In this work, we have described how thermal fluctuations and shear affect the stability of thin nanometric films described by the stochastic thin film equation. Finite element numerical solutions of the latter show that the role of the fluctuations is essentially to initiate perturbations of a characteristic size that is proportional to the thermal roughness of the interface. At later times, thermal fluctuations do not play a significant role in the dynamics of thin film rupture. By using a linear stability analysis, we give a rather simple prediction for the rupture time of a thin film as a function of initial height and temperature, in the absence of shear. When a shear force is introduced to a 2D system, rupture can be suppressed, resulting in the formation of a permanent travelling wave, while the initial size of the perturbations seems unaffected in the parameter range explored by the simulations. In the more physically realistic case of unidirectional shear in 3D films, however, rupture is not suppressed, as cross-shear perturbations grow unimpeded. Our results may explain why the reported rupture-inhibiting effect of shear has not been reported experimentally. This may be relevant in physical processes such as the thin air film formed below a droplet as it impacts on a surface, for which rupture is observed despite a high shear rate \cite{KolinskiRubenstein2012}.  Our simulations do indicate, however, that rupture can be delayed if the direction of shear varies sufficiently rapidly with time -- a situation of potential practical relevance for dewetting experiments and nanocoextrusion processes.

\begin{acknowledgments}
The authors acknowledge financial support from the European Union through the European Research Council under EMetBrown (ERCCoG-101039103) grant, from the Research Council of Norway through the program NANO2021 (project number 301138), and from the Agence Nationale de la Recherche under EMetBrown (ANR-21-ERCC-0010-01), Softer (ANR-21-CE06-0029) and Fricolas (ANR-21-CE06-0039) grants.
\end{acknowledgments}
\FloatBarrier
\appendix

 \section{Adaptive time step simulations}\label{sec:adaptiv}
\begin{figure}
	 \centering
	\begin{tikzpicture}
   	\draw (0, 0) node[inner sep=0] (fig) {\includegraphics[width=0.47\textwidth]{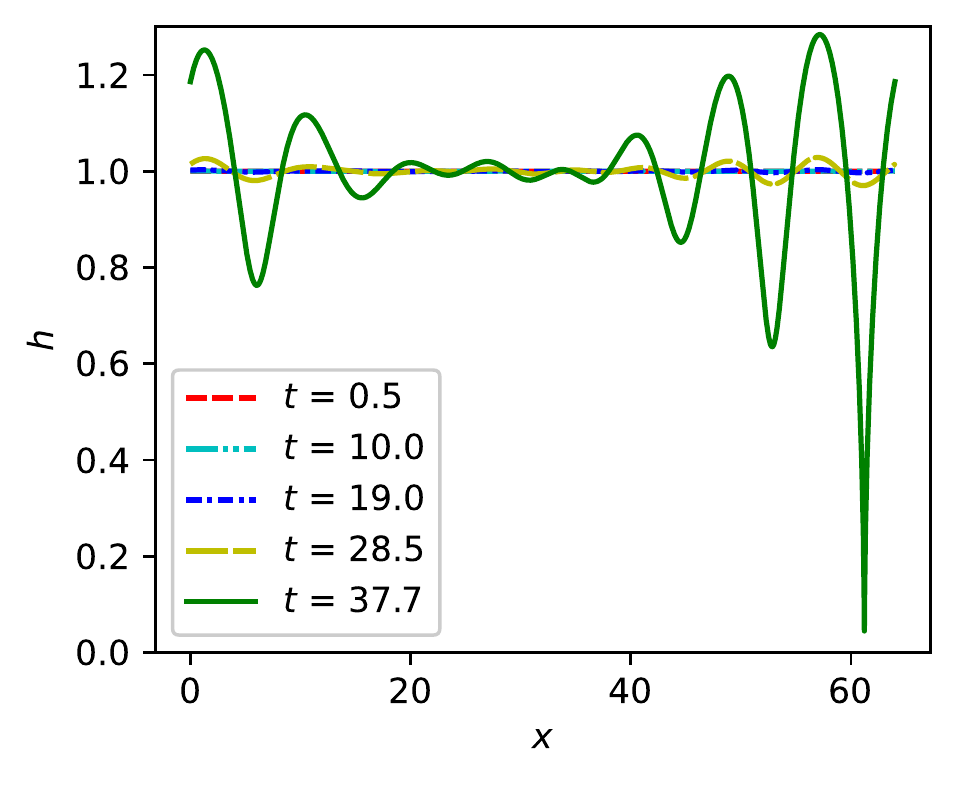}};
   	\node[] at (fig.north west){$(a)$};
 	\end{tikzpicture}
 	\begin{tikzpicture}
   	\draw (0, 0) node[inner sep=0] (fig) {\includegraphics[width=0.47\textwidth]{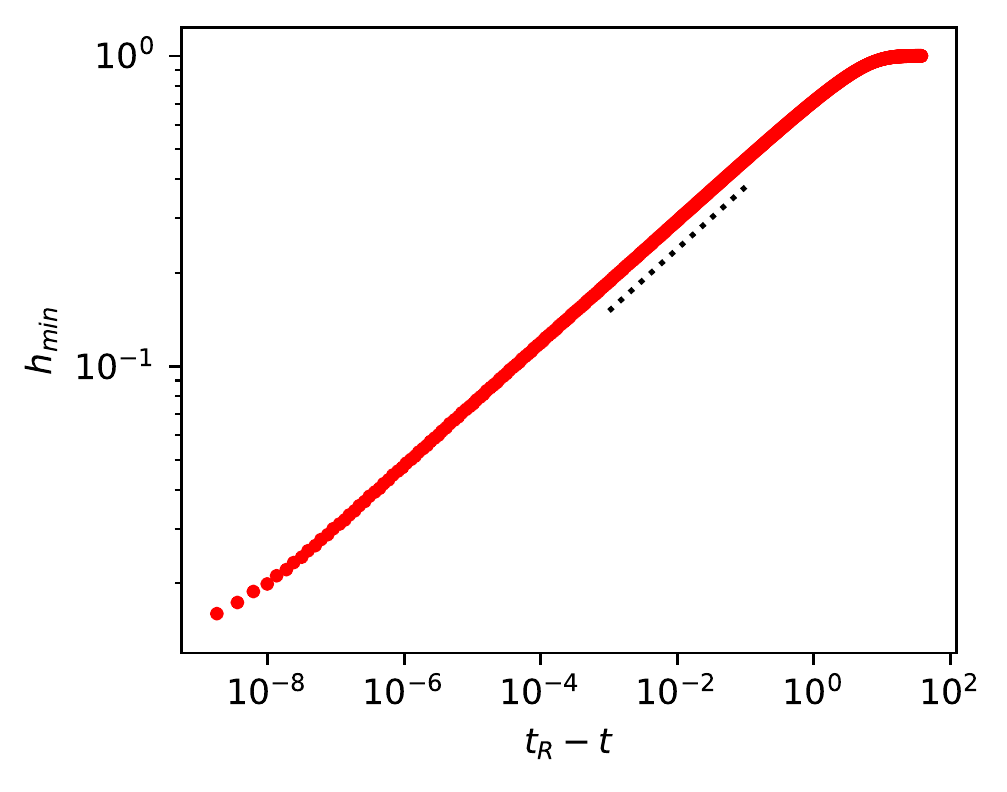}};
   	\node[] at (fig.north west){$(b)$};
 	\end{tikzpicture}
 	\begin{tikzpicture}
   	\draw (0, 0) node[inner sep=0] (fig) {\includegraphics[width=0.47\textwidth]{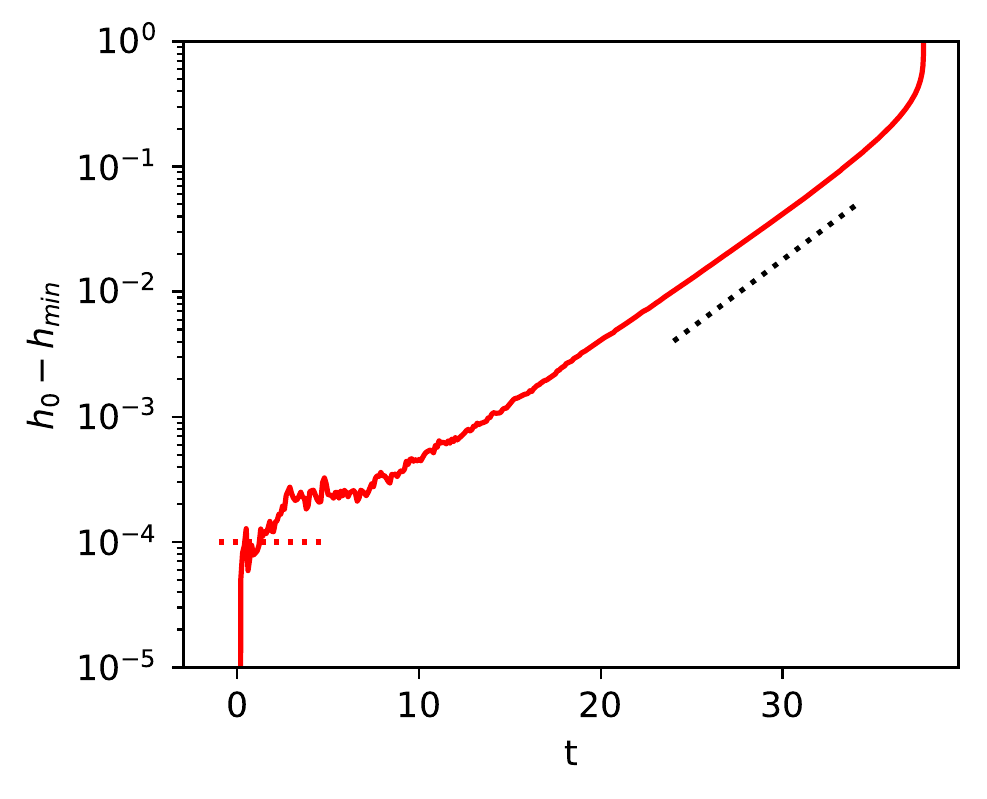}};
   	\node[] at (fig.north west){$(c)$};
 	\end{tikzpicture}

	\caption{Results of the simulations with an adaptive time step near rupture, for $Q=0.0001$, $B=0$, and $h_0=1$. $(a)$ Film profiles at various times. $(b)$ Minimum film height $h_{min}$ as a function of the reverse time $t_R-t$, in log-log scale. The $(t_R-t)^{1/5}$ power law \cite{ZhangLister1999} with arbitrary prefactor is indicated by the dotted black line. $(c)$ Amplitude of the height perturbation as a function of time.  The dotted black line line indicates exponential behaviour with a growth rate given by Eq. \eqref{eq:maxgrowth} and an arbitrary amplitude. The dotted red line indicates the $Q$ value.
	\label{fig:adaptiv}}.  
\end{figure}
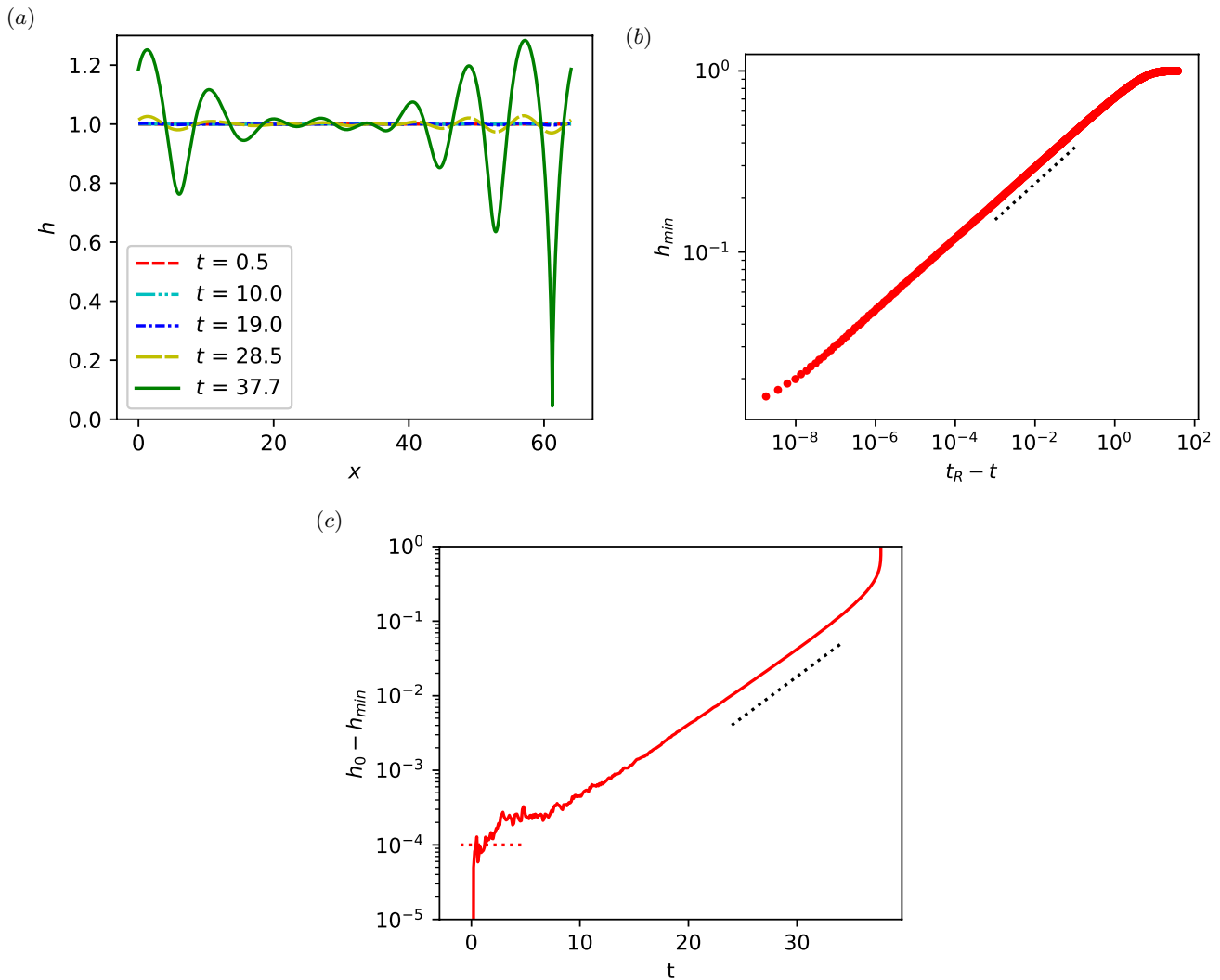

As described in section \ref{sec:Numerics}, we stop our simulations when the minimum height of the film reaches a threshold value of $h^*=0.33$, and define the time at which this occurs as the rupture time of the film, $t_R$. Our justification for this is that the late stage of rupture is an almost instantaneous event that follows the power law described by Zhang and Lister \cite{ZhangLister1999}. In order to confirm that this is indeed a good assumption, we have performed simulations as described in section \ref{sec:Numerics}, but with an adaptive time step that decreases as the film height decreases, as well as a reduced grid spacing of $\Delta x=0.001$ in order to resolve the flow details in this regime. Fig. \ref{fig:adaptiv} shows the results of such simulations. It is clear here that the late stage of rupture is accurately described by the aforementioned power law. The time elapsed between $h_{min}=0.33$ and $h_{min}=0.015$ is less than $0.1$, which is a negligible fraction of the rupture time $t_R$, for all the cases simulated in this article. The results when shear is included are similar.

\section{Thermal roughness of a flat interface with shear}\label{sec:shearroughness}

\begin{figure}[h]
	 \centering
	\begin{tikzpicture}
   	\draw (0, 0) node[inner sep=0] (fig) {\includegraphics[width=0.47\textwidth]{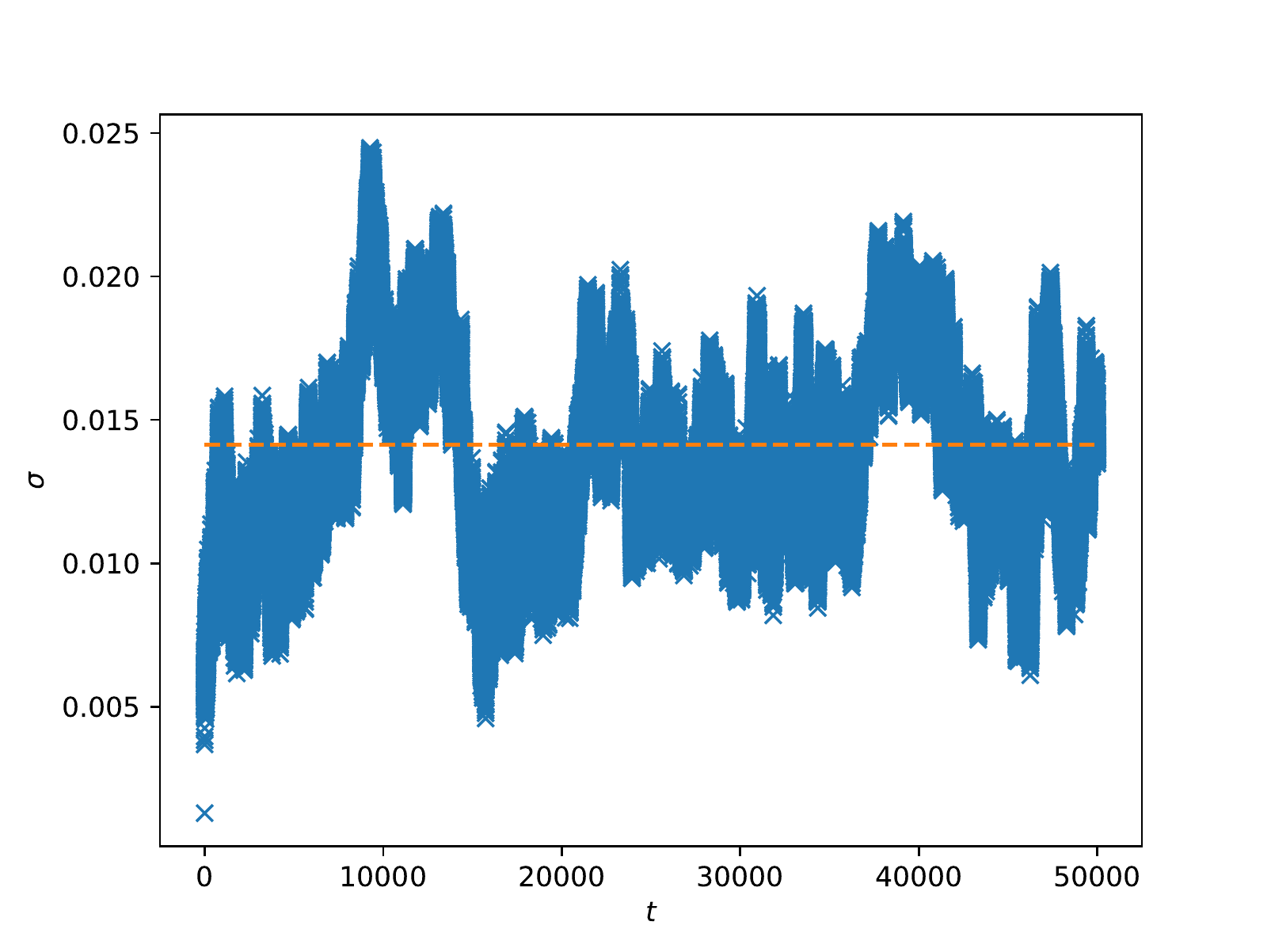}};
   	\node[] at (fig.north west){$(a)$};
 	\end{tikzpicture}
 	\begin{tikzpicture}
   	\draw (0, 0) node[inner sep=0] (fig) {\includegraphics[width=0.47\textwidth]{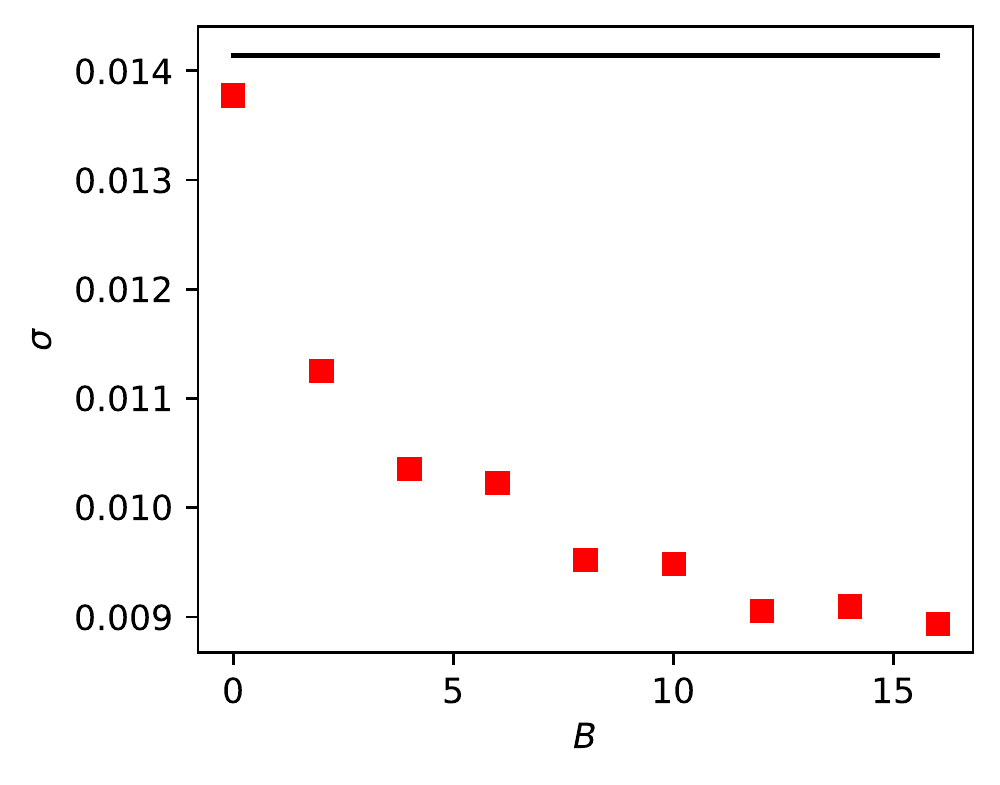}};
   	\node[] at (fig.north west){$(b)$};
 	\end{tikzpicture}
	\caption{Results of the simulations of a flat film without disjoining pressure with $Q=0.01$ and $h_0=1.0$. $(a)$ Instantaneous measured thermal roughness $\sigma$ as a function of time for the case of $B=0$. The dotted orange line shows the expected thermal roughness of $\sqrt{2}Q$. $(b)$ The thermal roughness as a function of the dimensionless shear force $B$. The black line shows the value of $\sqrt{2}Q$. \label{fig:roughness}}
\end{figure}

In section \ref{subsec:2Dresults}, we find that shear does not significantly change the initial size of the perturbations that lead to thin film rupture when thermal fluctuations are imposed. This is rather surprising in light of previous works which show that imposing a shear force significantly decreases the equilibrium thermal roughness of interfaces in other geometries \cite{ThiebaudBickel2015, Derks2006}. We thus simulate a 2D thin film subjected to thermal fluctuations and shear but no disjoining pressure, for which the interface fluctuates around its initial height $h_0$ indefinitely. We then compute the thermal roughness of the interface at each time step as $\sigma = (\sum_{n=1}^{N}(h-h_0)^2/N)^{1/2}$, where $n$ represents each of $N$ total gridpoints. In Fig. \ref{fig:roughness}$(a)$ we show how $\sigma$ varies in time when there is no shear. In both cases, $\sigma$ fluctuates around a mean value of around $Q$ as expected due to the definition of $Q$ in section \ref{sec:Equations}.

When we impose a shear $B \neq 0$ to the interface, we do indeed observe a decrease in the thermal roughness, as is shown in Fig. \ref{fig:roughness}$(b)$ . In fact, for $B >3$, $\sigma$ is reduced by more than half. Although consistent with what we expect from the literature, this result seems to conflict with the results in Fig. \ref{fig:2-D_shearprofiles}$(d)$ where the initial perturbation size is independent of $B$. Upon deeper reflection, however, it is not obvious that the equilibrium thermal roughness observed over time is the same as the size of instantaneous fluctuations. In our simulations of film rupture with both fluctuations and shear, it seems that perturbations to a flat film of size approximately $Q$ are created during the first few timesteps before the disjoining pressure or shear have any effect.  The sinusoidal perturbations that lead to rupture then form before the shear effect on thermal roughness has had any effect. It is thus important to distinguish between the time-averaged equilibrium effect of fluctuations and their instantaneous impact on the film.

\FloatBarrier
\bibliography{Bibliography_Vira}

\providecommand{\noopsort}[1]{}\providecommand{\singleletter}[1]{#1}%
\begin{thebibliography}{52}%
\makeatletter
\providecommand \@ifxundefined [1]{%
 \@ifx{#1\undefined}
}%
\providecommand \@ifnum [1]{%
 \ifnum #1\expandafter \@firstoftwo
 \else \expandafter \@secondoftwo
 \fi
}%
\providecommand \@ifx [1]{%
 \ifx #1\expandafter \@firstoftwo
 \else \expandafter \@secondoftwo
 \fi
}%
\providecommand \natexlab [1]{#1}%
\providecommand \enquote  [1]{``#1''}%
\providecommand \bibnamefont  [1]{#1}%
\providecommand \bibfnamefont [1]{#1}%
\providecommand \citenamefont [1]{#1}%
\providecommand \href@noop [0]{\@secondoftwo}%
\providecommand \href [0]{\begingroup \@sanitize@url \@href}%
\providecommand \@href[1]{\@@startlink{#1}\@@href}%
\providecommand \@@href[1]{\endgroup#1\@@endlink}%
\providecommand \@sanitize@url [0]{\catcode `\\12\catcode `\$12\catcode
  `\&12\catcode `\#12\catcode `\^12\catcode `\_12\catcode `\%12\relax}%
\providecommand \@@startlink[1]{}%
\providecommand \@@endlink[0]{}%
\providecommand \url  [0]{\begingroup\@sanitize@url \@url }%
\providecommand \@url [1]{\endgroup\@href {#1}{\urlprefix }}%
\providecommand \urlprefix  [0]{URL }%
\providecommand \Eprint [0]{\href }%
\providecommand \doibase [0]{https://doi.org/}%
\providecommand \selectlanguage [0]{\@gobble}%
\providecommand \bibinfo  [0]{\@secondoftwo}%
\providecommand \bibfield  [0]{\@secondoftwo}%
\providecommand \translation [1]{[#1]}%
\providecommand \BibitemOpen [0]{}%
\providecommand \bibitemStop [0]{}%
\providecommand \bibitemNoStop [0]{.\EOS\space}%
\providecommand \EOS [0]{\spacefactor3000\relax}%
\providecommand \BibitemShut  [1]{\csname bibitem#1\endcsname}%
\let\auto@bib@innerbib\@empty
\bibitem [{\citenamefont {Wong}\ \emph {et~al.}(1996)\citenamefont {Wong},
  \citenamefont {Fatt},\ and\ \citenamefont {Radke}}]{WongFattRadke1996}%
  \BibitemOpen
  \bibfield  {author} {\bibinfo {author} {\bibfnamefont {H.}~\bibnamefont
  {Wong}}, \bibinfo {author} {\bibfnamefont {I.}~\bibnamefont {Fatt}},\ and\
  \bibinfo {author} {\bibfnamefont {C.}~\bibnamefont {Radke}},\ }\bibfield
  {title} {\bibinfo {title} {Deposition and thinning of the human tear film},\
  }\href {https://doi.org/10.1006/jcis.1996.0595} {\bibfield  {journal}
  {\bibinfo  {journal} {Journal of Colloid and Interface Science}\ }\textbf
  {\bibinfo {volume} {184}},\ \bibinfo {pages} {44} (\bibinfo {year}
  {1996})}\BibitemShut {NoStop}%
\bibitem [{\citenamefont {{Chandran Suja}}\ \emph {et~al.}(2022)\citenamefont
  {{Chandran Suja}}, \citenamefont {Verma}, \citenamefont {Mossige},
  \citenamefont {Cui}, \citenamefont {Xia}, \citenamefont {Zhang},
  \citenamefont {Sinha}, \citenamefont {Joslin},\ and\ \citenamefont
  {Fuller}}]{SujaMossigeFuller2022}%
  \BibitemOpen
  \bibfield  {author} {\bibinfo {author} {\bibfnamefont {V.}~\bibnamefont
  {{Chandran Suja}}}, \bibinfo {author} {\bibfnamefont {A.}~\bibnamefont
  {Verma}}, \bibinfo {author} {\bibfnamefont {E.}~\bibnamefont {Mossige}},
  \bibinfo {author} {\bibfnamefont {K.}~\bibnamefont {Cui}}, \bibinfo {author}
  {\bibfnamefont {V.}~\bibnamefont {Xia}}, \bibinfo {author} {\bibfnamefont
  {Y.}~\bibnamefont {Zhang}}, \bibinfo {author} {\bibfnamefont
  {D.}~\bibnamefont {Sinha}}, \bibinfo {author} {\bibfnamefont
  {S.}~\bibnamefont {Joslin}},\ and\ \bibinfo {author} {\bibfnamefont
  {G.}~\bibnamefont {Fuller}},\ }\bibfield  {title} {\bibinfo {title}
  {Dewetting characteristics of contact lenses coated with wetting agents},\
  }\href@noop {} {\bibfield  {journal} {\bibinfo  {journal} {Journal of Colloid
  and Interface Science}\ }\textbf {\bibinfo {volume} {614}},\ \bibinfo {pages}
  {24} (\bibinfo {year} {2022})}\BibitemShut {NoStop}%
\bibitem [{\citenamefont {Li}\ \emph {et~al.}(2020)\citenamefont {Li},
  \citenamefont {Olah},\ and\ \citenamefont {Baer}}]{LiBaer2020}%
  \BibitemOpen
  \bibfield  {author} {\bibinfo {author} {\bibfnamefont {Z.}~\bibnamefont
  {Li}}, \bibinfo {author} {\bibfnamefont {A.}~\bibnamefont {Olah}},\ and\
  \bibinfo {author} {\bibfnamefont {E.}~\bibnamefont {Baer}},\ }\bibfield
  {title} {\bibinfo {title} {Micro- and nano- layered processing of new
  polymeric systems},\ }\href
  {https://doi.org/10.1016/j.progpolymsci.2020.101210} {\bibfield  {journal}
  {\bibinfo  {journal} {Progress in Polymer Science}\ }\textbf {\bibinfo
  {volume} {102}},\ \bibinfo {pages} {101210} (\bibinfo {year}
  {2020})}\BibitemShut {NoStop}%
\bibitem [{\citenamefont {Bironeau}\ \emph {et~al.}(2017)\citenamefont
  {Bironeau}, \citenamefont {Salez}, \citenamefont {Miquelard-Garnier},\ and\
  \citenamefont {Sollogoub}}]{BironeauSollogoub2017}%
  \BibitemOpen
  \bibfield  {author} {\bibinfo {author} {\bibfnamefont {A.}~\bibnamefont
  {Bironeau}}, \bibinfo {author} {\bibfnamefont {T.}~\bibnamefont {Salez}},
  \bibinfo {author} {\bibfnamefont {G.}~\bibnamefont {Miquelard-Garnier}},\
  and\ \bibinfo {author} {\bibfnamefont {C.}~\bibnamefont {Sollogoub}},\
  }\bibfield  {title} {\bibinfo {title} {Existence of a critical layer
  thickness in ps/pmma nanolayered films},\ }\href@noop {} {\bibfield
  {journal} {\bibinfo  {journal} {Macromolecules}\ }\textbf {\bibinfo {volume}
  {50}},\ \bibinfo {pages} {4064} (\bibinfo {year} {2017})}\BibitemShut
  {NoStop}%
\bibitem [{\citenamefont {Craster}\ and\ \citenamefont
  {Matar}(2009)}]{craster2009}%
  \BibitemOpen
  \bibfield  {author} {\bibinfo {author} {\bibfnamefont {R.}~\bibnamefont
  {Craster}}\ and\ \bibinfo {author} {\bibfnamefont {O.}~\bibnamefont
  {Matar}},\ }\bibfield  {title} {\bibinfo {title} {Dynamics and stability of
  thin liquid films},\ }\href {https://doi.org/10.1103/RevModPhys.81.1131}
  {\bibfield  {journal} {\bibinfo  {journal} {Reviews of Modern Physics}\
  }\textbf {\bibinfo {volume} {81}} (\bibinfo {year} {2009})}\BibitemShut
  {NoStop}%
\bibitem [{\citenamefont {Ferrell}\ and\ \citenamefont
  {Hansford}(2007)}]{FerrellHansford2007}%
  \BibitemOpen
  \bibfield  {author} {\bibinfo {author} {\bibfnamefont {N.}~\bibnamefont
  {Ferrell}}\ and\ \bibinfo {author} {\bibfnamefont {D.}~\bibnamefont
  {Hansford}},\ }\bibfield  {title} {\bibinfo {title} {Fabrication of micro-
  and nanoscale polymer structures by soft lithography and spin dewetting},\
  }\href@noop {} {\bibfield  {journal} {\bibinfo  {journal} {Macromolecular
  Rapid Communications}\ }\textbf {\bibinfo {volume} {28}},\ \bibinfo {pages}
  {966} (\bibinfo {year} {2007})}\BibitemShut {NoStop}%
\bibitem [{\citenamefont {Dhara}\ \emph {et~al.}(2018)\citenamefont {Dhara},
  \citenamefont {Bhandaru}, \citenamefont {Das},\ and\ \citenamefont
  {Mukherjee}}]{DharaMukherjee2018}%
  \BibitemOpen
  \bibfield  {author} {\bibinfo {author} {\bibfnamefont {P.}~\bibnamefont
  {Dhara}}, \bibinfo {author} {\bibfnamefont {N.}~\bibnamefont {Bhandaru}},
  \bibinfo {author} {\bibfnamefont {A.}~\bibnamefont {Das}},\ and\ \bibinfo
  {author} {\bibfnamefont {R.}~\bibnamefont {Mukherjee}},\ }\bibfield  {title}
  {\bibinfo {title} {Transition from spin dewetting to continuous film in spin
  coating of liquid crystal 5cb},\ }\href@noop {} {\bibfield  {journal}
  {\bibinfo  {journal} {Scientific Reports}\ }\textbf {\bibinfo {volume} {8}},\
  \bibinfo {pages} {7169} (\bibinfo {year} {2018})}\BibitemShut {NoStop}%
\bibitem [{\citenamefont {Vrij}(1966)}]{Vrij1966}%
  \BibitemOpen
  \bibfield  {author} {\bibinfo {author} {\bibfnamefont {A.}~\bibnamefont
  {Vrij}},\ }\bibfield  {title} {\bibinfo {title} {Possible mechanism for the
  spontaneous rupture of thin{,} free liquid films},\ }\href
  {https://doi.org/10.1039/DF9664200023} {\bibfield  {journal} {\bibinfo
  {journal} {Discuss. Faraday Soc.}\ }\textbf {\bibinfo {volume} {42}},\
  \bibinfo {pages} {23} (\bibinfo {year} {1966})}\BibitemShut {NoStop}%
\bibitem [{\citenamefont {Sheludko}(1967)}]{SHELUDKO1967}%
  \BibitemOpen
  \bibfield  {author} {\bibinfo {author} {\bibfnamefont {A.}~\bibnamefont
  {Sheludko}},\ }\bibfield  {title} {\bibinfo {title} {Thin liquid films},\
  }\href@noop {} {\bibfield  {journal} {\bibinfo  {journal} {Advances in
  Colloid and Interface Science}\ }\textbf {\bibinfo {volume} {1}},\ \bibinfo
  {pages} {391} (\bibinfo {year} {1967})}\BibitemShut {NoStop}%
\bibitem [{\citenamefont {Ruckenstein}\ and\ \citenamefont
  {Jain}(1974)}]{RuckensteinJain1974}%
  \BibitemOpen
  \bibfield  {author} {\bibinfo {author} {\bibfnamefont {E.}~\bibnamefont
  {Ruckenstein}}\ and\ \bibinfo {author} {\bibfnamefont {R.~K.}\ \bibnamefont
  {Jain}},\ }\bibfield  {title} {\bibinfo {title} {Spontaneous rupture of thin
  liquid films},\ }\href@noop {} {\bibfield  {journal} {\bibinfo  {journal} {J.
  Chem. Soc.{,} Faraday Trans. 2}\ }\textbf {\bibinfo {volume} {70}},\ \bibinfo
  {pages} {132} (\bibinfo {year} {1974})}\BibitemShut {NoStop}%
\bibitem [{\citenamefont {Williams}\ and\ \citenamefont
  {Davis}(1982)}]{WilliamsDavis1982}%
  \BibitemOpen
  \bibfield  {author} {\bibinfo {author} {\bibfnamefont {M.~B.}\ \bibnamefont
  {Williams}}\ and\ \bibinfo {author} {\bibfnamefont {S.~H.}\ \bibnamefont
  {Davis}},\ }\bibfield  {title} {\bibinfo {title} {Nonlinear theory of film
  rupture},\ }\href
  {https://doi.org/https://doi.org/10.1016/0021-9797(82)90415-5} {\bibfield
  {journal} {\bibinfo  {journal} {Journal of Colloid and Interface Science}\
  }\textbf {\bibinfo {volume} {90}},\ \bibinfo {pages} {220} (\bibinfo {year}
  {1982})}\BibitemShut {NoStop}%
\bibitem [{\citenamefont {Oron}\ \emph {et~al.}(1997)\citenamefont {Oron},
  \citenamefont {Davis},\ and\ \citenamefont {Bankoff}}]{OronBankoff1997}%
  \BibitemOpen
  \bibfield  {author} {\bibinfo {author} {\bibfnamefont {A.}~\bibnamefont
  {Oron}}, \bibinfo {author} {\bibfnamefont {S.}~\bibnamefont {Davis}},\ and\
  \bibinfo {author} {\bibfnamefont {S.}~\bibnamefont {Bankoff}},\ }\bibfield
  {title} {\bibinfo {title} {Long-scale evolution of thin liquid films},\
  }\href {https://doi.org/10.1103/RevModPhys.69.931} {\bibfield  {journal}
  {\bibinfo  {journal} {Reviews of Modern Physics}\ }\textbf {\bibinfo {volume}
  {69}},\ \bibinfo {pages} {931} (\bibinfo {year} {1997})}\BibitemShut
  {NoStop}%
\bibitem [{\citenamefont {Sharma}\ and\ \citenamefont
  {Ruckenstein}(1986)}]{SharmaRuck1986}%
  \BibitemOpen
  \bibfield  {author} {\bibinfo {author} {\bibfnamefont {A.}~\bibnamefont
  {Sharma}}\ and\ \bibinfo {author} {\bibfnamefont {E.}~\bibnamefont
  {Ruckenstein}},\ }\bibfield  {title} {\bibinfo {title} {An analytical
  nonlinear theory of thin film rupture and its application to wetting films},\
  }\href@noop {} {\bibfield  {journal} {\bibinfo  {journal} {Journal of Colloid
  and Interface Science}\ }\textbf {\bibinfo {volume} {113}},\ \bibinfo {pages}
  {456} (\bibinfo {year} {1986})}\BibitemShut {NoStop}%
\bibitem [{\citenamefont {Zhang}\ and\ \citenamefont
  {Lister}(1999)}]{ZhangLister1999}%
  \BibitemOpen
  \bibfield  {author} {\bibinfo {author} {\bibfnamefont {W.~W.}\ \bibnamefont
  {Zhang}}\ and\ \bibinfo {author} {\bibfnamefont {J.~R.}\ \bibnamefont
  {Lister}},\ }\bibfield  {title} {\bibinfo {title} {Similarity solutions for
  van der waals rupture of a thin film on a solid substrate},\ }\href@noop {}
  {\bibfield  {journal} {\bibinfo  {journal} {Physics of Fluids}\ }\textbf
  {\bibinfo {volume} {11}},\ \bibinfo {pages} {2454} (\bibinfo {year}
  {1999})}\BibitemShut {NoStop}%
\bibitem [{\citenamefont {Dallaston}\ \emph {et~al.}(2018)\citenamefont
  {Dallaston}, \citenamefont {Fontelos}, \citenamefont {Tseluiko},\ and\
  \citenamefont {Kalliadasis}}]{DallastonKalliadasis2018}%
  \BibitemOpen
  \bibfield  {author} {\bibinfo {author} {\bibfnamefont {M.~C.}\ \bibnamefont
  {Dallaston}}, \bibinfo {author} {\bibfnamefont {M.~A.}\ \bibnamefont
  {Fontelos}}, \bibinfo {author} {\bibfnamefont {D.}~\bibnamefont {Tseluiko}},\
  and\ \bibinfo {author} {\bibfnamefont {S.}~\bibnamefont {Kalliadasis}},\
  }\bibfield  {title} {\bibinfo {title} {Discrete self-similarity in
  interfacial hydrodynamics and the formation of iterated structures},\ }\href
  {https://doi.org/10.1103/PhysRevLett.120.034505} {\bibfield  {journal}
  {\bibinfo  {journal} {Phys. Rev. Lett.}\ }\textbf {\bibinfo {volume} {120}},\
  \bibinfo {pages} {034505} (\bibinfo {year} {2018})}\BibitemShut {NoStop}%
\bibitem [{\citenamefont {Carlson}\ and\ \citenamefont
  {Mahadevan}(2015)}]{carlson2015physfluids}%
  \BibitemOpen
  \bibfield  {author} {\bibinfo {author} {\bibfnamefont {A.}~\bibnamefont
  {Carlson}}\ and\ \bibinfo {author} {\bibfnamefont {L.}~\bibnamefont
  {Mahadevan}},\ }\bibfield  {title} {\bibinfo {title} {Protein mediated
  membrane adhesion},\ }\href@noop {} {\bibfield  {journal} {\bibinfo
  {journal} {Physics of Fluids}\ }\textbf {\bibinfo {volume} {27}},\ \bibinfo
  {pages} {051901} (\bibinfo {year} {2015})}\BibitemShut {NoStop}%
\bibitem [{\citenamefont {Becker}\ \emph {et~al.}(2003)\citenamefont {Becker},
  \citenamefont {Gr{\"u}n}, \citenamefont {Seemann}, \citenamefont {Mantz},
  \citenamefont {Jacobs}, \citenamefont {Mecke},\ and\ \citenamefont
  {Blossey}}]{BeckerJacobsMecke2003}%
  \BibitemOpen
  \bibfield  {author} {\bibinfo {author} {\bibfnamefont {J.}~\bibnamefont
  {Becker}}, \bibinfo {author} {\bibfnamefont {G.}~\bibnamefont {Gr{\"u}n}},
  \bibinfo {author} {\bibfnamefont {R.}~\bibnamefont {Seemann}}, \bibinfo
  {author} {\bibfnamefont {H.}~\bibnamefont {Mantz}}, \bibinfo {author}
  {\bibfnamefont {K.}~\bibnamefont {Jacobs}}, \bibinfo {author} {\bibfnamefont
  {K.~R.}\ \bibnamefont {Mecke}},\ and\ \bibinfo {author} {\bibfnamefont
  {R.}~\bibnamefont {Blossey}},\ }\bibfield  {title} {\bibinfo {title} {Complex
  dewetting scenarios captured by thin-film models},\ }\href@noop {} {\bibfield
   {journal} {\bibinfo  {journal} {Nature Materials}\ }\textbf {\bibinfo
  {volume} {2}},\ \bibinfo {pages} {59} (\bibinfo {year} {2003})}\BibitemShut
  {NoStop}%
\bibitem [{\citenamefont {Jacobs}\ \emph {et~al.}(2008)\citenamefont {Jacobs},
  \citenamefont {Seemann},\ and\ \citenamefont
  {Herminghaus}}]{JacobsSeemannHerminghaus2008}%
  \BibitemOpen
  \bibfield  {author} {\bibinfo {author} {\bibfnamefont {K.}~\bibnamefont
  {Jacobs}}, \bibinfo {author} {\bibfnamefont {R.}~\bibnamefont {Seemann}},\
  and\ \bibinfo {author} {\bibfnamefont {S.}~\bibnamefont {Herminghaus}},\
  }\bibfield  {title} {\bibinfo {title} {Stability and dewetting of thin liquid
  films},\ }\href {https://doi.org/10.1142/9789812818829_0010} {\bibfield
  {journal} {\bibinfo  {journal} {Polymer Thin Films}\ }\textbf {\bibinfo
  {volume} {1}} (\bibinfo {year} {2008})}\BibitemShut {NoStop}%
\bibitem [{\citenamefont {Davidovitch}\ \emph {et~al.}(2005)\citenamefont
  {Davidovitch}, \citenamefont {Moro},\ and\ \citenamefont
  {Stone}}]{davidovitch2005}%
  \BibitemOpen
  \bibfield  {author} {\bibinfo {author} {\bibfnamefont {B.}~\bibnamefont
  {Davidovitch}}, \bibinfo {author} {\bibfnamefont {E.}~\bibnamefont {Moro}},\
  and\ \bibinfo {author} {\bibfnamefont {H.~A.}\ \bibnamefont {Stone}},\
  }\bibfield  {title} {\bibinfo {title} {Spreading of viscous fluid drops on a
  solid substrate assisted by thermal fluctuations},\ }\href
  {https://doi.org/10.1103/PhysRevLett.95.244505} {\bibfield  {journal}
  {\bibinfo  {journal} {Phys. Rev. Lett.}\ }\textbf {\bibinfo {volume} {95}},\
  \bibinfo {pages} {244505} (\bibinfo {year} {2005})}\BibitemShut {NoStop}%
\bibitem [{\citenamefont {Grün}\ \emph {et~al.}(2006)\citenamefont {Grün},
  \citenamefont {Mecke},\ and\ \citenamefont {Rauscher}}]{grun2006}%
  \BibitemOpen
  \bibfield  {author} {\bibinfo {author} {\bibfnamefont {G.}~\bibnamefont
  {Grün}}, \bibinfo {author} {\bibfnamefont {K.}~\bibnamefont {Mecke}},\ and\
  \bibinfo {author} {\bibfnamefont {M.}~\bibnamefont {Rauscher}},\ }\bibfield
  {title} {\bibinfo {title} {Thin-film flow influenced by thermal noise},\
  }\href {https://doi.org/10.1007/s10955-006-9028-8} {\bibfield  {journal}
  {\bibinfo  {journal} {Journal of Statistical Physics}\ }\textbf {\bibinfo
  {volume} {122}},\ \bibinfo {pages} {1261} (\bibinfo {year}
  {2006})}\BibitemShut {NoStop}%
\bibitem [{\citenamefont {Carlson}(2018)}]{carlson_2018}%
  \BibitemOpen
  \bibfield  {author} {\bibinfo {author} {\bibfnamefont {A.}~\bibnamefont
  {Carlson}},\ }\bibfield  {title} {\bibinfo {title} {Fluctuation assisted
  spreading of a fluid filled elastic blister},\ }\href
  {https://doi.org/10.1017/jfm.2018.288} {\bibfield  {journal} {\bibinfo
  {journal} {Journal of Fluid Mechanics}\ }\textbf {\bibinfo {volume} {846}},\
  \bibinfo {pages} {1076–1087} (\bibinfo {year} {2018})}\BibitemShut
  {NoStop}%
\bibitem [{\citenamefont {Pedersen}\ \emph {et~al.}(2019)\citenamefont
  {Pedersen}, \citenamefont {Niven}, \citenamefont {Salez}, \citenamefont
  {Dalnoki-Veress},\ and\ \citenamefont
  {Carlson}}]{pedersen_niven_salez_dalnoki-veress_carlson_2019}%
  \BibitemOpen
  \bibfield  {author} {\bibinfo {author} {\bibfnamefont {C.}~\bibnamefont
  {Pedersen}}, \bibinfo {author} {\bibfnamefont {J.~F.}\ \bibnamefont {Niven}},
  \bibinfo {author} {\bibfnamefont {T.}~\bibnamefont {Salez}}, \bibinfo
  {author} {\bibfnamefont {K.}~\bibnamefont {Dalnoki-Veress}},\ and\ \bibinfo
  {author} {\bibfnamefont {A.}~\bibnamefont {Carlson}},\ }\bibfield  {title}
  {\bibinfo {title} {Asymptotic regimes in elastohydrodynamic and stochastic
  leveling on a viscous film},\ }\href@noop {} {\bibfield  {journal} {\bibinfo
  {journal} {Physical Review Fluids}\ }\textbf {\bibinfo {volume} {4}}
  (\bibinfo {year} {2019})}\BibitemShut {NoStop}%
\bibitem [{\citenamefont {Marbach}\ \emph {et~al.}(2018)\citenamefont
  {Marbach}, \citenamefont {Dean},\ and\ \citenamefont
  {Bocquet}}]{marbach_dean_bocquet_2018}%
  \BibitemOpen
  \bibfield  {author} {\bibinfo {author} {\bibfnamefont {S.}~\bibnamefont
  {Marbach}}, \bibinfo {author} {\bibfnamefont {D.~S.}\ \bibnamefont {Dean}},\
  and\ \bibinfo {author} {\bibfnamefont {L.}~\bibnamefont {Bocquet}},\
  }\bibfield  {title} {\bibinfo {title} {Transport and dispersion across
  wiggling nanopores},\ }\href {https://doi.org/10.1038/s41567-018-0239-0}
  {\bibfield  {journal} {\bibinfo  {journal} {Nature Physics}\ }\textbf
  {\bibinfo {volume} {14}},\ \bibinfo {pages} {1108–1113} (\bibinfo {year}
  {2018})}\BibitemShut {NoStop}%
\bibitem [{\citenamefont {Diez}\ \emph {et~al.}(2016)\citenamefont {Diez},
  \citenamefont {González},\ and\ \citenamefont
  {Fernandez}}]{DiezFernandez2016}%
  \BibitemOpen
  \bibfield  {author} {\bibinfo {author} {\bibfnamefont {J.}~\bibnamefont
  {Diez}}, \bibinfo {author} {\bibfnamefont {A.}~\bibnamefont {González}},\
  and\ \bibinfo {author} {\bibfnamefont {R.}~\bibnamefont {Fernandez}},\
  }\bibfield  {title} {\bibinfo {title} {Metallic-thin-film instability with
  spatially correlated thermal noise},\ }\href
  {https://doi.org/10.1103/PhysRevE.93.013120} {\bibfield  {journal} {\bibinfo
  {journal} {Physical Review E}\ }\textbf {\bibinfo {volume} {93}} (\bibinfo
  {year} {2016})}\BibitemShut {NoStop}%
\bibitem [{\citenamefont {Nesic}\ \emph {et~al.}(2015)\citenamefont {Nesic},
  \citenamefont {Cuerno}, \citenamefont {Moro},\ and\ \citenamefont
  {Kondic}}]{NesicKondic2015}%
  \BibitemOpen
  \bibfield  {author} {\bibinfo {author} {\bibfnamefont {S.}~\bibnamefont
  {Nesic}}, \bibinfo {author} {\bibfnamefont {R.}~\bibnamefont {Cuerno}},
  \bibinfo {author} {\bibfnamefont {E.}~\bibnamefont {Moro}},\ and\ \bibinfo
  {author} {\bibfnamefont {L.}~\bibnamefont {Kondic}},\ }\bibfield  {title}
  {\bibinfo {title} {Fully nonlinear dynamics of stochastic thin-film
  dewetting},\ }\href {https://doi.org/10.1103/PhysRevE.92.061002} {\bibfield
  {journal} {\bibinfo  {journal} {Phys. Rev. E}\ }\textbf {\bibinfo {volume}
  {92}},\ \bibinfo {pages} {061002} (\bibinfo {year} {2015})}\BibitemShut
  {NoStop}%
\bibitem [{\citenamefont {Mecke}\ and\ \citenamefont
  {Rauscher}(2005)}]{MeckeRauscher_2005}%
  \BibitemOpen
  \bibfield  {author} {\bibinfo {author} {\bibfnamefont {K.}~\bibnamefont
  {Mecke}}\ and\ \bibinfo {author} {\bibfnamefont {M.}~\bibnamefont
  {Rauscher}},\ }\bibfield  {title} {\bibinfo {title} {On thermal fluctuations
  in thin film flow},\ }\href@noop {} {\bibfield  {journal} {\bibinfo
  {journal} {Journal of Physics: Condensed Matter}\ }\textbf {\bibinfo {volume}
  {17}},\ \bibinfo {pages} {S3515} (\bibinfo {year} {2005})}\BibitemShut
  {NoStop}%
\bibitem [{\citenamefont {Dur{\'a}n-Olivencia}\ \emph
  {et~al.}(2019)\citenamefont {Dur{\'a}n-Olivencia}, \citenamefont {Gvalani},
  \citenamefont {Kalliadasis},\ and\ \citenamefont
  {Pavliotis}}]{duran-olivenciakalliadasis}%
  \BibitemOpen
  \bibfield  {author} {\bibinfo {author} {\bibfnamefont {M.~A.}\ \bibnamefont
  {Dur{\'a}n-Olivencia}}, \bibinfo {author} {\bibfnamefont {R.~S.}\
  \bibnamefont {Gvalani}}, \bibinfo {author} {\bibfnamefont {S.}~\bibnamefont
  {Kalliadasis}},\ and\ \bibinfo {author} {\bibfnamefont {G.~A.}\ \bibnamefont
  {Pavliotis}},\ }\bibfield  {title} {\bibinfo {title} {Instability, rupture
  and fluctuations in thin liquid films: Theory and computations},\ }\href
  {https://doi.org/10.1007/s10955-018-2200-0} {\bibfield  {journal} {\bibinfo
  {journal} {Journal of Statistical Physics}\ }\textbf {\bibinfo {volume}
  {174}},\ \bibinfo {pages} {579} (\bibinfo {year} {2019})}\BibitemShut
  {NoStop}%
\bibitem [{\citenamefont {Shah}\ \emph {et~al.}(2019)\citenamefont {Shah},
  \citenamefont {Steijn}, \citenamefont {Kleijn},\ and\ \citenamefont
  {Kreutzer}}]{ShahKreutzer}%
  \BibitemOpen
  \bibfield  {author} {\bibinfo {author} {\bibfnamefont {M.}~\bibnamefont
  {Shah}}, \bibinfo {author} {\bibfnamefont {V.}~\bibnamefont {Steijn}},
  \bibinfo {author} {\bibfnamefont {C.}~\bibnamefont {Kleijn}},\ and\ \bibinfo
  {author} {\bibfnamefont {M.}~\bibnamefont {Kreutzer}},\ }\bibfield  {title}
  {\bibinfo {title} {Thermal fluctuations in capillary thinning of thin liquid
  films},\ }\href {https://doi.org/10.1017/jfm.2019.595} {\bibfield  {journal}
  {\bibinfo  {journal} {Journal of Fluid Mechanics}\ }\textbf {\bibinfo
  {volume} {876}},\ \bibinfo {pages} {1090} (\bibinfo {year}
  {2019})}\BibitemShut {NoStop}%
\bibitem [{\citenamefont {Zhao}\ \emph {et~al.}(2022)\citenamefont {Zhao},
  \citenamefont {Liu}, \citenamefont {Lockerby},\ and\ \citenamefont
  {Sprittles}}]{Zhaosprittles2022}%
  \BibitemOpen
  \bibfield  {author} {\bibinfo {author} {\bibfnamefont {C.}~\bibnamefont
  {Zhao}}, \bibinfo {author} {\bibfnamefont {J.}~\bibnamefont {Liu}}, \bibinfo
  {author} {\bibfnamefont {D.~A.}\ \bibnamefont {Lockerby}},\ and\ \bibinfo
  {author} {\bibfnamefont {J.~E.}\ \bibnamefont {Sprittles}},\ }\bibfield
  {title} {\bibinfo {title} {Fluctuation-driven dynamics in nanoscale thin-film
  flows: Physical insights from numerical investigations},\ }\href
  {https://doi.org/10.1103/PhysRevFluids.7.024203} {\bibfield  {journal}
  {\bibinfo  {journal} {Phys. Rev. Fluids}\ }\textbf {\bibinfo {volume} {7}},\
  \bibinfo {pages} {024203} (\bibinfo {year} {2022})}\BibitemShut {NoStop}%
\bibitem [{\citenamefont {Derks}\ \emph {et~al.}(2006)\citenamefont {Derks},
  \citenamefont {Aarts}, \citenamefont {Bonn}, \citenamefont {Lekkerkerker},\
  and\ \citenamefont {Imhof}}]{Derks2006}%
  \BibitemOpen
  \bibfield  {author} {\bibinfo {author} {\bibfnamefont {D.}~\bibnamefont
  {Derks}}, \bibinfo {author} {\bibfnamefont {D.~G. A.~L.}\ \bibnamefont
  {Aarts}}, \bibinfo {author} {\bibfnamefont {D.}~\bibnamefont {Bonn}},
  \bibinfo {author} {\bibfnamefont {H.~N.~W.}\ \bibnamefont {Lekkerkerker}},\
  and\ \bibinfo {author} {\bibfnamefont {A.}~\bibnamefont {Imhof}},\ }\bibfield
   {title} {\bibinfo {title} {Suppression of thermally excited capillary waves
  by shear flow},\ }\href {https://doi.org/10.1103/PhysRevLett.97.038301}
  {\bibfield  {journal} {\bibinfo  {journal} {Phys. Rev. Lett.}\ }\textbf
  {\bibinfo {volume} {97}},\ \bibinfo {pages} {038301} (\bibinfo {year}
  {2006})}\BibitemShut {NoStop}%
\bibitem [{\citenamefont {Thiébaud}\ and\ \citenamefont
  {Bickel}(2010)}]{ThiebaudBickel2015}%
  \BibitemOpen
  \bibfield  {author} {\bibinfo {author} {\bibfnamefont {M.}~\bibnamefont
  {Thiébaud}}\ and\ \bibinfo {author} {\bibfnamefont {T.}~\bibnamefont
  {Bickel}},\ }\bibfield  {title} {\bibinfo {title} {Nonequilibrium
  fluctuations of an interface under shear},\ }\href
  {https://doi.org/10.1103/PhysRevE.81.031602} {\bibfield  {journal} {\bibinfo
  {journal} {Physical review. E, Statistical, nonlinear, and soft matter
  physics}\ }\textbf {\bibinfo {volume} {81}},\ \bibinfo {pages} {031602}
  (\bibinfo {year} {2010})}\BibitemShut {NoStop}%
\bibitem [{\citenamefont {Bresson}\ \emph {et~al.}(2017)\citenamefont
  {Bresson}, \citenamefont {Brun}, \citenamefont {Buet}, \citenamefont {Chen},
  \citenamefont {Ciccotti}, \citenamefont {G\^ateau}, \citenamefont {Jasion},
  \citenamefont {Petrovich}, \citenamefont {Poletti}, \citenamefont
  {Richardson}, \citenamefont {Sandoghchi}, \citenamefont {Tessier},
  \citenamefont {Tyukodi},\ and\ \citenamefont {Vandembroucq}}]{Bresson2017}%
  \BibitemOpen
  \bibfield  {author} {\bibinfo {author} {\bibfnamefont {B.}~\bibnamefont
  {Bresson}}, \bibinfo {author} {\bibfnamefont {C.}~\bibnamefont {Brun}},
  \bibinfo {author} {\bibfnamefont {X.}~\bibnamefont {Buet}}, \bibinfo {author}
  {\bibfnamefont {Y.}~\bibnamefont {Chen}}, \bibinfo {author} {\bibfnamefont
  {M.}~\bibnamefont {Ciccotti}}, \bibinfo {author} {\bibfnamefont
  {J.}~\bibnamefont {G\^ateau}}, \bibinfo {author} {\bibfnamefont
  {G.}~\bibnamefont {Jasion}}, \bibinfo {author} {\bibfnamefont {M.~N.}\
  \bibnamefont {Petrovich}}, \bibinfo {author} {\bibfnamefont {F.}~\bibnamefont
  {Poletti}}, \bibinfo {author} {\bibfnamefont {D.~J.}\ \bibnamefont
  {Richardson}}, \bibinfo {author} {\bibfnamefont {S.~R.}\ \bibnamefont
  {Sandoghchi}}, \bibinfo {author} {\bibfnamefont {G.}~\bibnamefont {Tessier}},
  \bibinfo {author} {\bibfnamefont {B.}~\bibnamefont {Tyukodi}},\ and\ \bibinfo
  {author} {\bibfnamefont {D.}~\bibnamefont {Vandembroucq}},\ }\bibfield
  {title} {\bibinfo {title} {Anisotropic superattenuation of capillary waves on
  driven glass interfaces},\ }\href
  {https://doi.org/10.1103/PhysRevLett.119.235501} {\bibfield  {journal}
  {\bibinfo  {journal} {Phys. Rev. Lett.}\ }\textbf {\bibinfo {volume} {119}},\
  \bibinfo {pages} {235501} (\bibinfo {year} {2017})}\BibitemShut {NoStop}%
\bibitem [{\citenamefont {Kalpathy}\ \emph {et~al.}(2010)\citenamefont
  {Kalpathy}, \citenamefont {Francis},\ and\ \citenamefont
  {Kumar}}]{KalpathyKumar2010}%
  \BibitemOpen
  \bibfield  {author} {\bibinfo {author} {\bibfnamefont {S.}~\bibnamefont
  {Kalpathy}}, \bibinfo {author} {\bibfnamefont {L.}~\bibnamefont {Francis}},\
  and\ \bibinfo {author} {\bibfnamefont {S.}~\bibnamefont {Kumar}},\ }\bibfield
   {title} {\bibinfo {title} {Shear-induced suppression of rupture in two-layer
  thin liquid films},\ }\href {https://doi.org/10.1016/j.jcis.2010.04.028}
  {\bibfield  {journal} {\bibinfo  {journal} {Journal of colloid and interface
  science}\ }\textbf {\bibinfo {volume} {348}},\ \bibinfo {pages} {271}
  (\bibinfo {year} {2010})}\BibitemShut {NoStop}%
\bibitem [{\citenamefont {Davis}\ \emph {et~al.}(2010)\citenamefont {Davis},
  \citenamefont {Gratton},\ and\ \citenamefont
  {Davis}}]{davis_gratton_davis_2010}%
  \BibitemOpen
  \bibfield  {author} {\bibinfo {author} {\bibfnamefont {M.~J.}\ \bibnamefont
  {Davis}}, \bibinfo {author} {\bibfnamefont {M.~B.}\ \bibnamefont {Gratton}},\
  and\ \bibinfo {author} {\bibfnamefont {S.~H.}\ \bibnamefont {Davis}},\
  }\bibfield  {title} {\bibinfo {title} {Suppressing van der waals driven
  rupture through shear},\ }\href {https://doi.org/10.1017/S002211201000323X}
  {\bibfield  {journal} {\bibinfo  {journal} {Journal of Fluid Mechanics}\
  }\textbf {\bibinfo {volume} {661}},\ \bibinfo {pages} {522–539} (\bibinfo
  {year} {2010})}\BibitemShut {NoStop}%
\bibitem [{\citenamefont {Kadri}\ \emph {et~al.}(2021)\citenamefont {Kadri},
  \citenamefont {Peixinho}, \citenamefont {Salez}, \citenamefont
  {Miquelard-Garnier},\ and\ \citenamefont {Sollogoub}}]{Kadri2021}%
  \BibitemOpen
  \bibfield  {author} {\bibinfo {author} {\bibfnamefont {K.}~\bibnamefont
  {Kadri}}, \bibinfo {author} {\bibfnamefont {J.}~\bibnamefont {Peixinho}},
  \bibinfo {author} {\bibfnamefont {T.}~\bibnamefont {Salez}}, \bibinfo
  {author} {\bibfnamefont {G.}~\bibnamefont {Miquelard-Garnier}},\ and\
  \bibinfo {author} {\bibfnamefont {C.}~\bibnamefont {Sollogoub}},\ }\bibfield
  {title} {\bibinfo {title} {Dewetting of a thin polymer film under shear},\
  }\href {https://doi.org/https://doi.org/10.1016/j.polymer.2021.124283}
  {\bibfield  {journal} {\bibinfo  {journal} {Polymer}\ }\textbf {\bibinfo
  {volume} {235}},\ \bibinfo {pages} {124283} (\bibinfo {year}
  {2021})}\BibitemShut {NoStop}%
\bibitem [{\citenamefont {Dmochowska}\ \emph {et~al.}(2022)\citenamefont
  {Dmochowska}, \citenamefont {Peixinho}, \citenamefont {Sollogoub},\ and\
  \citenamefont {Miquelard-Garnier}}]{DmochowskaMiquelard-Garnier2022}%
  \BibitemOpen
  \bibfield  {author} {\bibinfo {author} {\bibfnamefont {A.}~\bibnamefont
  {Dmochowska}}, \bibinfo {author} {\bibfnamefont {J.}~\bibnamefont
  {Peixinho}}, \bibinfo {author} {\bibfnamefont {C.}~\bibnamefont
  {Sollogoub}},\ and\ \bibinfo {author} {\bibfnamefont {G.}~\bibnamefont
  {Miquelard-Garnier}},\ }\bibfield  {title} {\bibinfo {title} {Dewetting
  dynamics of sheared thin polymer films: An experimental study},\ }\href@noop
  {} {\bibfield  {journal} {\bibinfo  {journal} {ACS Macro Letters}\ }\textbf
  {\bibinfo {volume} {11}},\ \bibinfo {pages} {422} (\bibinfo {year}
  {2022})}\BibitemShut {NoStop}%
\bibitem [{\citenamefont {Batchelor}(2000)}]{batchelor_2000}%
  \BibitemOpen
  \bibfield  {author} {\bibinfo {author} {\bibfnamefont {G.~K.}\ \bibnamefont
  {Batchelor}},\ }\href {https://doi.org/10.1017/CBO9780511800955} {\emph
  {\bibinfo {title} {An Introduction to Fluid Dynamics}}},\ Cambridge
  Mathematical Library\ (\bibinfo  {publisher} {Cambridge University Press},\
  \bibinfo {year} {2000})\BibitemShut {NoStop}%
\bibitem [{\citenamefont {Israelachvili}(2011)}]{ISRAELACHVILI2011}%
  \BibitemOpen
  \bibfield  {author} {\bibinfo {author} {\bibfnamefont {J.~N.}\ \bibnamefont
  {Israelachvili}},\ }\href@noop {} {\emph {\bibinfo {title} {Intermolecular
  and Surface Forces}}},\ \bibinfo {edition} {third edition}\ ed.\ (\bibinfo
  {publisher} {Academic Press},\ \bibinfo {address} {San Diego},\ \bibinfo
  {year} {2011})\ pp.\ \bibinfo {pages} {253--289}\BibitemShut {NoStop}%
\bibitem [{\citenamefont {Logg}\ \emph {et~al.}(2012)\citenamefont {Logg},
  \citenamefont {Mardal}, \citenamefont {Wells} \emph
  {et~al.}}]{LoggMardalEtAl2012}%
  \BibitemOpen
  \bibfield  {author} {\bibinfo {author} {\bibfnamefont {A.}~\bibnamefont
  {Logg}}, \bibinfo {author} {\bibfnamefont {K.-A.}\ \bibnamefont {Mardal}},
  \bibinfo {author} {\bibfnamefont {G.~N.}\ \bibnamefont {Wells}}, \emph
  {et~al.},\ }\href {https://doi.org/10.1007/978-3-642-23099-8} {\emph
  {\bibinfo {title} {Automated Solution of Differential Equations by the Finite
  Element Method}}},\ edited by\ \bibinfo {editor} {\bibfnamefont
  {A.}~\bibnamefont {Logg}}, \bibinfo {editor} {\bibfnamefont {K.-A.}\
  \bibnamefont {Mardal}},\ and\ \bibinfo {editor} {\bibfnamefont {G.~N.}\
  \bibnamefont {Wells}}\ (\bibinfo  {publisher} {Springer},\ \bibinfo {year}
  {2012})\BibitemShut {NoStop}%
\bibitem [{\citenamefont {Oliphant}(2006)}]{OliphantNumpy2006}%
  \BibitemOpen
  \bibfield  {author} {\bibinfo {author} {\bibfnamefont {T.}~\bibnamefont
  {Oliphant}},\ }\href@noop {} {\emph {\bibinfo {title} {Guide to NumPy}}}\
  (\bibinfo {year} {2006})\BibitemShut {NoStop}%
\bibitem [{\citenamefont {Moreno-Boza}\ \emph {et~al.}(2020)\citenamefont
  {Moreno-Boza}, \citenamefont {Mart{\'\i}nez-Calvo},\ and\ \citenamefont
  {Sevilla}}]{morenoBozaMartinezCalvoSevilla2020}%
  \BibitemOpen
  \bibfield  {author} {\bibinfo {author} {\bibfnamefont {D.}~\bibnamefont
  {Moreno-Boza}}, \bibinfo {author} {\bibfnamefont {A.}~\bibnamefont
  {Mart{\'\i}nez-Calvo}},\ and\ \bibinfo {author} {\bibfnamefont
  {A.}~\bibnamefont {Sevilla}},\ }\bibfield  {title} {\bibinfo {title} {Stokes
  theory of thin-film rupture},\ }\href@noop {} {\bibfield  {journal} {\bibinfo
   {journal} {Physical Review Fluids}\ }\textbf {\bibinfo {volume} {5}},\
  \bibinfo {pages} {014002} (\bibinfo {year} {2020})}\BibitemShut {NoStop}%
\bibitem [{\citenamefont {Gholami}\ \emph {et~al.}(2020)\citenamefont
  {Gholami}, \citenamefont {Pakzad},\ and\ \citenamefont
  {Behzadfar}}]{Gholami20}%
  \BibitemOpen
  \bibfield  {author} {\bibinfo {author} {\bibfnamefont {F.}~\bibnamefont
  {Gholami}}, \bibinfo {author} {\bibfnamefont {L.}~\bibnamefont {Pakzad}},\
  and\ \bibinfo {author} {\bibfnamefont {E.}~\bibnamefont {Behzadfar}},\
  }\bibfield  {title} {\bibinfo {title} {Morphological, interfacial and
  rheological properties in multilayer polymers: A review},\ }\href@noop {}
  {\bibfield  {journal} {\bibinfo  {journal} {Polymer}\ }\textbf {\bibinfo
  {volume} {208}},\ \bibinfo {pages} {122950} (\bibinfo {year}
  {2020})}\BibitemShut {NoStop}%
\bibitem [{\citenamefont {Rijal}\ \emph {et~al.}(2022)\citenamefont {Rijal},
  \citenamefont {Delbreilh}, \citenamefont {Sollogoub}, \citenamefont {Baer},\
  and\ \citenamefont {Saiter-Fourcin}}]{Rijal22}%
  \BibitemOpen
  \bibfield  {author} {\bibinfo {author} {\bibfnamefont {B.}~\bibnamefont
  {Rijal}}, \bibinfo {author} {\bibfnamefont {L.}~\bibnamefont {Delbreilh}},
  \bibinfo {author} {\bibfnamefont {C.}~\bibnamefont {Sollogoub}}, \bibinfo
  {author} {\bibfnamefont {E.}~\bibnamefont {Baer}},\ and\ \bibinfo {author}
  {\bibfnamefont {A.}~\bibnamefont {Saiter-Fourcin}},\ }\bibfield  {title}
  {\bibinfo {title} {Multiscale analysis of segmental relaxation in {PC}/{PET}g
  multilayers: Evidence of immiscible nanodroplets},\ }\href@noop {} {\bibfield
   {journal} {\bibinfo  {journal} {Macromolecules}\ }\textbf {\bibinfo {volume}
  {55}},\ \bibinfo {pages} {6562} (\bibinfo {year} {2022})}\BibitemShut
  {NoStop}%
\bibitem [{\citenamefont {Lozay}\ \emph {et~al.}(2021)\citenamefont {Lozay},
  \citenamefont {Beuguel}, \citenamefont {Follain}, \citenamefont {Lebrun},
  \citenamefont {Guinault}, \citenamefont {Miquelard-Garnier}, \citenamefont
  {Tenc{\'e}-Girault}, \citenamefont {Sollogoub}, \citenamefont {Dargent},\
  and\ \citenamefont {Marais}}]{Lozay21}%
  \BibitemOpen
  \bibfield  {author} {\bibinfo {author} {\bibfnamefont {Q.}~\bibnamefont
  {Lozay}}, \bibinfo {author} {\bibfnamefont {Q.}~\bibnamefont {Beuguel}},
  \bibinfo {author} {\bibfnamefont {N.}~\bibnamefont {Follain}}, \bibinfo
  {author} {\bibfnamefont {L.}~\bibnamefont {Lebrun}}, \bibinfo {author}
  {\bibfnamefont {A.}~\bibnamefont {Guinault}}, \bibinfo {author}
  {\bibfnamefont {G.}~\bibnamefont {Miquelard-Garnier}}, \bibinfo {author}
  {\bibfnamefont {S.}~\bibnamefont {Tenc{\'e}-Girault}}, \bibinfo {author}
  {\bibfnamefont {C.}~\bibnamefont {Sollogoub}}, \bibinfo {author}
  {\bibfnamefont {E.}~\bibnamefont {Dargent}},\ and\ \bibinfo {author}
  {\bibfnamefont {S.}~\bibnamefont {Marais}},\ }\bibfield  {title} {\bibinfo
  {title} {Structural and barrier properties of compatibilized pe/pa6
  multinanolayer films},\ }\href@noop {} {\bibfield  {journal} {\bibinfo
  {journal} {Membranes}\ }\textbf {\bibinfo {volume} {11}},\ \bibinfo {pages}
  {75} (\bibinfo {year} {2021})}\BibitemShut {NoStop}%
\bibitem [{\citenamefont {Nassar}\ \emph {et~al.}(2018)\citenamefont {Nassar},
  \citenamefont {Domenek}, \citenamefont {Guinault}, \citenamefont {Stoclet},
  \citenamefont {Delpouve},\ and\ \citenamefont {Sollogoub}}]{Nassar18}%
  \BibitemOpen
  \bibfield  {author} {\bibinfo {author} {\bibfnamefont {S.~F.}\ \bibnamefont
  {Nassar}}, \bibinfo {author} {\bibfnamefont {S.}~\bibnamefont {Domenek}},
  \bibinfo {author} {\bibfnamefont {A.}~\bibnamefont {Guinault}}, \bibinfo
  {author} {\bibfnamefont {G.}~\bibnamefont {Stoclet}}, \bibinfo {author}
  {\bibfnamefont {N.}~\bibnamefont {Delpouve}},\ and\ \bibinfo {author}
  {\bibfnamefont {C.}~\bibnamefont {Sollogoub}},\ }\bibfield  {title} {\bibinfo
  {title} {Structural and dynamic heterogeneity in the amorphous phase of poly
  (l, l-lactide) confined at the nanoscale by the coextrusion process},\
  }\href@noop {} {\bibfield  {journal} {\bibinfo  {journal} {Macromolecules}\
  }\textbf {\bibinfo {volume} {51}},\ \bibinfo {pages} {128} (\bibinfo {year}
  {2018})}\BibitemShut {NoStop}%
\bibitem [{\citenamefont {Wu}(1970)}]{Wu70}%
  \BibitemOpen
  \bibfield  {author} {\bibinfo {author} {\bibfnamefont {S.}~\bibnamefont
  {Wu}},\ }\bibfield  {title} {\bibinfo {title} {Surface and interfacial
  tensions of polymer melts. ii. poly (methyl methacrylate), poly (n-butyl
  methacrylate), and polystyrene},\ }\href@noop {} {\bibfield  {journal}
  {\bibinfo  {journal} {The Journal of Physical Chemistry}\ }\textbf {\bibinfo
  {volume} {74}},\ \bibinfo {pages} {632} (\bibinfo {year} {1970})}\BibitemShut
  {NoStop}%
\bibitem [{\citenamefont {de~Silva}\ \emph {et~al.}(2012)\citenamefont
  {de~Silva}, \citenamefont {Cousin}, \citenamefont {Wildes}, \citenamefont
  {Geoghegan},\ and\ \citenamefont {Sferrazza}}]{deSilva12}%
  \BibitemOpen
  \bibfield  {author} {\bibinfo {author} {\bibfnamefont {J.~P.}\ \bibnamefont
  {de~Silva}}, \bibinfo {author} {\bibfnamefont {F.}~\bibnamefont {Cousin}},
  \bibinfo {author} {\bibfnamefont {A.~R.}\ \bibnamefont {Wildes}}, \bibinfo
  {author} {\bibfnamefont {M.}~\bibnamefont {Geoghegan}},\ and\ \bibinfo
  {author} {\bibfnamefont {M.}~\bibnamefont {Sferrazza}},\ }\bibfield  {title}
  {\bibinfo {title} {Symmetric and asymmetric instability of buried polymer
  interfaces},\ }\href@noop {} {\bibfield  {journal} {\bibinfo  {journal}
  {Physical Review E}\ }\textbf {\bibinfo {volume} {86}},\ \bibinfo {pages}
  {032801} (\bibinfo {year} {2012})}\BibitemShut {NoStop}%
\bibitem [{\citenamefont {Bironeau}(2016)}]{Bironeau2016}%
  \BibitemOpen
  \bibfield  {author} {\bibinfo {author} {\bibfnamefont {A.}~\bibnamefont
  {Bironeau}},\ }\emph {\bibinfo {title} {Films multinanocouches de polymères
  amorphes coextrudés : élaboration, caractérisation et stabilité des
  nanocouches}},\ \href@noop {} {Ph.D. thesis},\ \bibinfo  {school} {ENSAM}
  (\bibinfo {year} {2016})\BibitemShut {NoStop}%
\bibitem [{\citenamefont {Lenz}\ and\ \citenamefont {Kumar}(2007)}]{Lenz07}%
  \BibitemOpen
  \bibfield  {author} {\bibinfo {author} {\bibfnamefont {R.~D.}\ \bibnamefont
  {Lenz}}\ and\ \bibinfo {author} {\bibfnamefont {S.}~\bibnamefont {Kumar}},\
  }\bibfield  {title} {\bibinfo {title} {Instability of confined thin liquid
  film trilayers},\ }\href@noop {} {\bibfield  {journal} {\bibinfo  {journal}
  {Journal of Colloid and Interface Science}\ }\textbf {\bibinfo {volume}
  {316}},\ \bibinfo {pages} {660} (\bibinfo {year} {2007})}\BibitemShut
  {NoStop}%
\bibitem [{\citenamefont {Zhu}\ \emph {et~al.}(2016)\citenamefont {Zhu},
  \citenamefont {Bironeau}, \citenamefont {Restagno}, \citenamefont
  {Sollogoub},\ and\ \citenamefont
  {Miquelard-Garnier}}]{ZHUMiquelardGarnier2016}%
  \BibitemOpen
  \bibfield  {author} {\bibinfo {author} {\bibfnamefont {Y.}~\bibnamefont
  {Zhu}}, \bibinfo {author} {\bibfnamefont {A.}~\bibnamefont {Bironeau}},
  \bibinfo {author} {\bibfnamefont {F.}~\bibnamefont {Restagno}}, \bibinfo
  {author} {\bibfnamefont {C.}~\bibnamefont {Sollogoub}},\ and\ \bibinfo
  {author} {\bibfnamefont {G.}~\bibnamefont {Miquelard-Garnier}},\ }\bibfield
  {title} {\bibinfo {title} {Kinetics of thin polymer film rupture: Model
  experiments for a better understanding of layer breakups in the multilayer
  coextrusion process},\ }\href@noop {} {\bibfield  {journal} {\bibinfo
  {journal} {Polymer}\ }\textbf {\bibinfo {volume} {90}},\ \bibinfo {pages}
  {156} (\bibinfo {year} {2016})}\BibitemShut {NoStop}%
\bibitem [{\citenamefont {Chebil}\ \emph {et~al.}(2018)\citenamefont {Chebil},
  \citenamefont {McGraw}, \citenamefont {Salez}, \citenamefont {Sollogoub},\
  and\ \citenamefont {Miquelard-Garnier}}]{Chebil18}%
  \BibitemOpen
  \bibfield  {author} {\bibinfo {author} {\bibfnamefont {M.~S.}\ \bibnamefont
  {Chebil}}, \bibinfo {author} {\bibfnamefont {J.~D.}\ \bibnamefont {McGraw}},
  \bibinfo {author} {\bibfnamefont {T.}~\bibnamefont {Salez}}, \bibinfo
  {author} {\bibfnamefont {C.}~\bibnamefont {Sollogoub}},\ and\ \bibinfo
  {author} {\bibfnamefont {G.}~\bibnamefont {Miquelard-Garnier}},\ }\bibfield
  {title} {\bibinfo {title} {Influence of outer-layer finite-size effects on
  the dewetting dynamics of a thin polymer film embedded in an immiscible
  matrix},\ }\href@noop {} {\bibfield  {journal} {\bibinfo  {journal} {Soft
  matter}\ }\textbf {\bibinfo {volume} {14}},\ \bibinfo {pages} {6256}
  (\bibinfo {year} {2018})}\BibitemShut {NoStop}%
\bibitem [{\citenamefont {Kolinski}\ \emph {et~al.}(2012)\citenamefont
  {Kolinski}, \citenamefont {Rubinstein}, \citenamefont {Mandre}, \citenamefont
  {Brenner}, \citenamefont {Weitz},\ and\ \citenamefont
  {Mahadevan}}]{KolinskiRubenstein2012}%
  \BibitemOpen
  \bibfield  {author} {\bibinfo {author} {\bibfnamefont {J.~M.}\ \bibnamefont
  {Kolinski}}, \bibinfo {author} {\bibfnamefont {S.~M.}\ \bibnamefont
  {Rubinstein}}, \bibinfo {author} {\bibfnamefont {S.}~\bibnamefont {Mandre}},
  \bibinfo {author} {\bibfnamefont {M.~P.}\ \bibnamefont {Brenner}}, \bibinfo
  {author} {\bibfnamefont {D.~A.}\ \bibnamefont {Weitz}},\ and\ \bibinfo
  {author} {\bibfnamefont {L.}~\bibnamefont {Mahadevan}},\ }\bibfield  {title}
  {\bibinfo {title} {Skating on a film of air: Drops impacting on a surface},\
  }\href {https://doi.org/10.1103/PhysRevLett.108.074503} {\bibfield  {journal}
  {\bibinfo  {journal} {Phys. Rev. Lett.}\ }\textbf {\bibinfo {volume} {108}},\
  \bibinfo {pages} {074503} (\bibinfo {year} {2012})}\BibitemShut {NoStop}%
\end{thebibliography}%

\end{document}